\documentclass[9pt,twoside,twocolumn,english]{IEEEtran}

\usepackage[T1]{fontenc}
\usepackage[latin9]{inputenc}
\usepackage{xcolor}
\usepackage{amsmath}
\usepackage{amssymb}
\usepackage{graphicx}
\PassOptionsToPackage{normalem}{ulem}
\usepackage{ulem}

\makeatletter

\newcommand{\lyxdot}{.}

\providecolor{lyxadded}{rgb}{0,0,1}
\providecolor{lyxdeleted}{rgb}{1,0,0}

\DeclareRobustCommand{\lyxdeleted}[3]{{\color{lyxdeleted}\lyxsout{#3}}}
\DeclareRobustCommand{\lyxsout}[1]{\ifx\\#1\else\sout{#1}\fi}

\let\oldforeign@language\foreign@language
\DeclareRobustCommand{\foreign@language}[1]{%
  \lowercase{\oldforeign@language{#1}}}

\usepackage{cite}
\usepackage[caption=false,font=footnotesize]{subfig}

\PassOptionsToPackage{normalem}{ulem}
\usepackage{ulem}
    \providecolor{lyxadded}{rgb}{0,0,1}
    \providecolor{lyxdeleted}{rgb}{1,1,1}
    
    \DeclareRobustCommand{\lyxdeleted}[3]{{\color{lyxdeleted}\sout{#3}}}
\renewcommand{\lyxdeleted}[3]{}

\@ifundefined{showcaptionsetup}{}{%
 \PassOptionsToPackage{caption=false}{subfig}}
\usepackage{subfig}
\makeatother

\usepackage{babel}
\begin{document}
\twocolumn
\title{General Framework for Array Noise Analysis and Noise Performance of
a Two-Element Interferometer with a Mutual-Coupling Canceler}
\author{Leonid Belostotski, \IEEEmembership{Senior Member,~IEEE}, Adrian
Sutinjo, \IEEEmembership{Senior Member,~IEEE}, Ravi Subrahmanyan,
Soumyajit~Mandal,~\IEEEmembership{Senior Member,~IEEE}, and Arjuna
Madanayake, \IEEEmembership{Member,~IEEE} \thanks{Manuscript received on XXXX XX, 202X.  This work was supported by the University of Calgary, the Natural Sciences and Engineering Research Council of Canada, Canada Reasearch Chair Program, and CMC Microsystems.}\thanks{L. Belostotski is with the Department of Electrical and Software Engineering, University of Calgary (e-mail: lbelosto@ucalgary.ca).}\thanks{A. Sutinjo is with the School of Elecrical Engineering, Computer and Math Science, Curtin University.}\thanks{R. Subrahmanyan is with CSIRO Astronomy \& Space Science, Kensington WA, Australia.}\thanks{S. Mandal is with the  Instrumentation Division at Brookhaven National Laboratory, New York.}\thanks{A. Madanayake is with the Department of Electrical and Computer Engineering, Florida International University.}}
\markboth{IEEE Transactions on Antennas and Propagation,\  Vol. X, No. XX,
XXX 202X}{Belostotski, \MakeLowercase{\textit{et al.}: \MakeUppercase{General
Framework for Array Noise Analysis and Noise Performance of a Two-Element
Interferometer ...}}}
\maketitle
\begin{abstract}
This paper investigates the noise performance of a two-element phased
array and interferometer containing a recently introduced self-interference
canceler, which in the context of this work acts as a mutual-coupling
canceler. To this end, a general framework is proposed to permit noise
analysis of this network and a large variety of other networks. The
framework-based numerical analysis for a two-element phased array
shows that the addition of the canceler significantly increases the
beam-equivalent noise temperature. For a two-element interferometer
used in cosmology, this increase in noise temperature is still acceptable
as the sky noise temperature in the 20-to-200-MHz band is high. When
used in an interferometer, the canceler provides the ability to null
mutual coherence at the interferometer output. The ability to provide
matching to reduce the sensitivity of the null in mutual coherence
to the phase of the $\boldsymbol{90^{\circ}}$ hybrids in the canceler
is discussed.  
\end{abstract}


\thispagestyle{empty}
\def\contentsname{Contents}
\def\listfigurename{List of Figures}
\def\listtablename{List of Tables}
\def\refname{References}
\def\indexname{Index}  
\def\figurename{Fig.}  
\def\tablename{TABLE}   
\def\partname{Part}  
\def\appendixname{Appendix}    
\def\abstractname{Abstract}
\setlength{\parskip}{0pt}
\bstctlcite{IEEEexample:BSTcontrol}

\section{\label{sec:Introduction}Introduction}

\IEEEPARstart{A}{ntenna} arrays are increasingly used in modern communication
and remote-sensing systems, scientific instruments, military equipment,
and biomedical devices \cite{Zwart2002,Krieger2008,Heath2008,Dewdney2009,Zhang2010,Bookner2010,Warnick2010b,Reynolds2011,Rebeiz2013,Rappaport6G,HolographicMIMO}.
 In comparison to single-antenna transceivers, the design of antenna-array
transceivers is challenging due to electromagnetic coupling between
antennas \cite{Maaskant2020,Chippendale2014}. However, antenna arrays
offer new opportunities, such as, for example, beam steering and noise
shaping \cite{wang2018delta,Rebeiz2021}. Two-element arrays form
an important subset of arrays that are both used on their own and
also as tools for studying the performance of larger arrays on a smaller
and tractable scale.

This paper investigates the impact of a recently introduced wideband
self-interference canceler, which was originally aimed at full-duplex
multiple-input multiple-output (MIMO) wireless systems \cite{Madanayake2021}
and is similar to an earlier work in \cite{Beyers2015}. The cancellation
is based on replica antennas that are located in isolation within
an electromagnetically shielded absorptive box. The replica antennas
mimic mutual coupling and scattering of the main antennas, thus allowing
passive microwave-based subtraction of coupled components from the
signals of interest \cite{Madanayake2021}. While an inspection of
the canceler network suggests that this replica-antenna-based canceler
is expected to increase the beam-equivalent noise temperature, $T_{\text{rec}}$,
of such an array, the canceler provides the possibility of decoupling
the antennas and thus has potential uses in radio cosmology.

Radio cosmology instruments typically rely on interferometric measurements
to discern 10s to 100s mK spectrally rich perturbations in the Cosmic
Microwave Background of $3\,\text{K}$, that is itself much weaker
than the Galactic noise of $\sim10^{2}$ to $10^{4}\,\text{K}$. Fortunately,
while Galactic signals are strong, their smooth spectra permit their
separation from cosmological signals. A single-antenna radiometer
has already been used for cosmology experiments between 50 and 100
MHz \cite{Monsalve2017}. Interferometric instruments are now being
investigated \cite{Venumadhav_2016} to verify the outcomes of \cite{Monsalve2017}.

A cosmology interferometer is expected to consist of two closely spaced
antennas connected to receiver circuitry followed by a correlator.
The close antenna spacing ($<0.4\lambda$, where $\lambda$ is the
wavelength) realizes correlation for isotropic noise of the surrounding
medium \cite{SutinjoURI2020} that is subsequently detected by the
correlator for measuring the noise temperature of a uniform sky, $T_{\text{sky}}$.
Design equations for such a system, without the decoupling network,
were recently presented in \cite{SutinjoURI2020}, where it was shown
that the cross-correlation does indeed contain the desired signal,
particularly when the antenna spacing is minimized. However, the mutual
coupling of closely spaced antennas results in the noise from each
receiver propagating to the output through multiple paths, thereby
contaminating the output by the undesirable mutual coherence. To reduce
this contamination, a decoupling network may be employed \cite{Andersen1976,Wang2004,Hein2006,Kossiavas2006,Hein2008,Warnick2007}.
While isolators form an intuitive implementation of such a network,
their bandwidths are limited to approximately 10\% in the 20-to-200-MHz
frequency range. Instead, a wideband decoupling network stemming from
the work in \cite{Madanayake2021} is considered here. As is shown,
this network is able to decouple antennas and thereby null the mutual-coupling-induced
correlation in two-element interferometer-based cosmology instruments.

In Section \ref{sec:Antenna-array-with-Canceler}, we derive key equations
that form a general framework for describing the noise performance
of an $M$-element array. This framework is used in the rest of the
paper, which is divided into two sections: Section \ref{sec:Receiver-noise-temperature:}
discusses the beam-equivalent noise temperature, $T_{\text{rec}}$,
of the two-element array, and Section \ref{sec:Output-cross-correlation}
investigates the mutual coherence of the system. Finally, Section
\ref{sec:Discussions} discusses our results, and Section \ref{sec:Conclusion}
concludes the paper.

\section{\label{sec:Antenna-array-with-Canceler}Antenna Array with the Canceler}

\begin{figure}
\begin{centering}
\includegraphics[width=1\columnwidth]{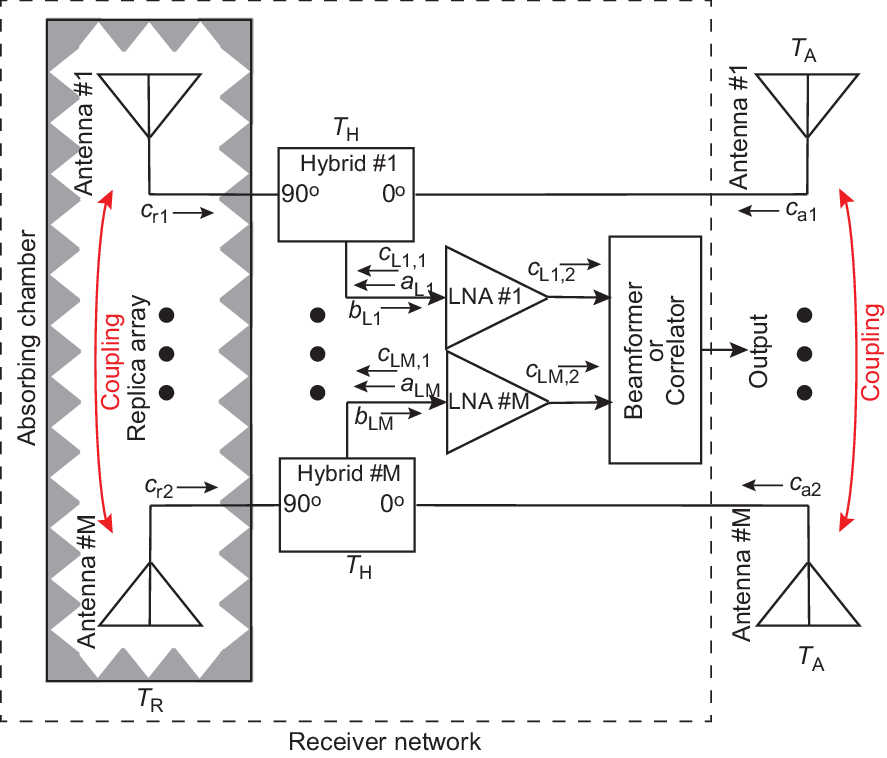}
\par\end{centering}
\caption{\label{fig:Block-diagram}Block diagram of an $M$-element array with
the mutual-coupling canceler implemented with a replica array and
$90^{\circ}$ hybrids.}
\end{figure}

The noise of antenna arrays directly connected to receivers has been
investigated in the past by many groups \cite{Belostotski2009_TAP,Belostotski_TAP2015,Maaskant2008,Sutinjo_Ung_20202,Findeklee2011}.
These investigations have adopted different approaches to deriving
expressions for noise analysis but they are all specific to their
particular systems and are not applicable to the network discussed
here. Therefore, we start by developing a general framework that can
be adopted for any other network while also avoiding the need for
manual error-prone signal-flow-graph analysis. We use this framework
to analyze the noise behavior of the system shown in Fig. \ref{fig:Block-diagram}.
This block diagram is a receiver-only implementation of the system
based on \cite{Madanayake2021}. The main array consists of the $M$-antenna
array, labeled as Antenna \#1 to \#$M$, and referred to as ``antenna
array'' in the rest of the paper. The replica array is nominally
identical to the main array but placed in an absorbing chamber. Unlike
\cite{Madanayake2021}, two nominally identical $90^{\circ}$ hybrids
are used. In addition to preserving the cancellation property analyzed
in \cite{Madanayake2021}, these hybrids have two advantages: the
reflection coefficients as seen by the LNAs only depend on the hybrid
and do not depend on the antenna array, and maintaining similar performance
between two $90^{\circ}$ hybrids is easier than between the splitter
and $180^{\circ}$ hybrid used in the system in \cite{Madanayake2021}.
The receiver network, as identified in the figure, consists of the
replica array, hybrids, LNAs, and a beamformer or correlator. The
noise of the complete system is modeled by noise waves $\mathbf{c}$
emanating out of input and output ports of all sub-components. In
a conventional implementation, i.e. without the replica array and
the hybrids, the noise emanating from the LNA inputs, such as $c_{\text{L1},1}$
to $c_{\text{LM},1}$, couples to adjacent antennas and propagates
to the beamformer. Such coupling causes non-zero cross correlation
\cite{SutinjoURI2020}. With the replica array in place, $c_{\text{L}i,1}$,
where $i=1\ldots M$, propagate to the beamformer through two paths
while experiencing a $180^{\circ}$ phase shift, thus providing the
possibility of cancellation. While cancellation of coupled $c_{\text{L}i,1}$
is therefore possible, the additional absorbing chamber and hybrids
undoubtedly contribute additional noise.

\subsection{\label{subsec:Key-Equations}Key Equations}

A set of linear equations that relate traveling waves at the ports
of the complete network is
\begin{equation}
\mathbf{b}=\mathbf{S}\left(\mathbf{a}+\mathbf{a}_{\text{s}}\right)+\mathbf{c}\label{eq:b-Sa}
\end{equation}
with 
\begin{equation}
\mathbf{S}=\left[\begin{array}{cc}
\mathbf{S}_{\text{PASS}} & \mathbf{0}\\
\mathbf{0} & \mathbf{S}_{\text{ACT}}
\end{array}\right]\label{eq:S-matrix-general}
\end{equation}
where in general $\mathbf{S}_{\text{PASS}}$ and $\mathbf{S}_{\text{ACT}}$
are S-parameter matrices of the passive and active sub-circuits, respectively;
$\mathbf{a}$ is a vector of all incident waves; $\mathbf{b}$ is
a vector of all reflected waves; $\mathbf{a}_{\text{s}}$ is a vector
of generated waves of sources connected to the network ports; and
the noise waves are represented by the vector $\mathbf{c}$.

The connections between all internally interconnected ports in the
system are identified by $\mathbf{K}$, where $\mathbf{a}=\mathbf{K}\mathbf{b}$,
such that \eqref{eq:b-Sa} becomes
\begin{equation}
\mathbf{b}=\mathbf{S}\left(\mathbf{K}\mathbf{b}+\mathbf{a}_{\text{s}}\right)+\mathbf{c}\label{eq:b-bs-c}
\end{equation}
and forms the main expression that is used throughout this work to
determine the noise and gain of the system.

Whereas \eqref{eq:b-Sa} to \eqref{eq:b-bs-c} are general, for the
particular system in Fig. \ref{fig:Block-diagram}, the elements of
\eqref{eq:b-Sa} and \eqref{eq:S-matrix-general} are 
\begin{equation}
\mathbf{S}_{\text{PASS}}=\left[\begin{array}{ccccc}
\mathbf{S}_{\text{A}_{M\ensuremath{\times M}}}\\
 & \mathbf{S}_{\text{R}_{M\ensuremath{\times M}}} &  & \mathbf{0}\\
 &  & \mathbf{S}_{\text{H1\ensuremath{_{3\times3}}}} & \mathbf{}\\
 & \mathbf{0} &  & \ddots\\
 &  &  &  & \mathbf{S}_{\text{HM}_{3\ensuremath{\times3}}}
\end{array}\right],\label{eq:S-pass}
\end{equation}
\begin{equation}
\mathbf{S}_{\text{ACT}}=\left[\begin{array}{ccc}
\mathbf{S}_{\text{L1}{}_{2\times2}} &  & \mathbf{0}\\
 & \ddots\\
\mathbf{0} &  & \mathbf{S}_{\text{LM}_{2\times2}}
\end{array}\right],\label{eq:S-act}
\end{equation}
\begin{equation}
\begin{array}{c}
\mathbf{a}=\left[\begin{array}{cccc}
\mathbf{a}_{\text{A}} & \mathbf{a}_{\text{R}} & \mathbf{a}_{\text{H}} & \mathbf{a}_{\text{L}}\end{array}\right]_{1\times7M}^{T},\\
\mathbf{b}=\left[\begin{array}{cccc}
\mathbf{b}_{\text{A}} & \mathbf{b}_{\text{R}} & \mathbf{b}_{\text{H}} & \mathbf{b}_{\text{L}}\end{array}\right]_{1\times7M}^{T},\\
\mathbf{c}=\left[\begin{array}{cccccc}
\mathbf{c}_{\text{A}} & \mathbf{c}_{\text{R}} & \mathbf{c}_{\text{H1}} & \mathbf{c}_{\text{H2}} & \mathbf{c}_{\text{L1}} & \mathbf{c}_{\text{L2}}\end{array}\right]_{1\times7M}^{T}\\
=\left[\begin{array}{cc}
\mathbf{c}_{\text{PASS}} & \mathbf{c}_{\text{ACT}}\end{array}\right]^{T},
\end{array}
\end{equation}
where subscripts A, R, H1(2), and L1(2) identify the antenna
array, replica array, hybrids \#1(\#2), and LNAs \#1(\#2), respectively.
The vector $\mathbf{c}$ consists of individual noise contributions
from each passive and active parts of the antenna network. For a two-antenna
system, $\mathbf{c}_{\text{A}}=\left[\begin{array}{cc}
c_{\text{a1}} & c_{\text{a2}}\end{array}\right]$, $\mathbf{c}_{\text{R}}=\left[\begin{array}{cc}
c_{\text{r1}} & c_{\text{r2}}\end{array}\right]$, $\mathbf{c}_{\text{H1}}=\left[\begin{array}{ccc}
c_{\text{h1,1}} & c_{\text{h1,2}} & c_{\text{h1,3}}\end{array}\right]$, $\mathbf{c}_{\text{H2}}=\left[\begin{array}{ccc}
c_{\text{h2,1}} & c_{\text{h2,2}} & c_{\text{h2,3}}\end{array}\right]$, $\mathbf{c}_{\text{L1}}=\left[\begin{array}{cc}
c_{\text{L1,1}} & c_{\text{L1,2}}\end{array}\right]$, and $\mathbf{c}_{\text{L2}}=\left[\begin{array}{cc}
c_{\text{L2,1}} & c_{\text{L2,2}}\end{array}\right]$. If required, matrix $\mathbf{S}$ in \eqref{eq:S-pass} can also
include S-parameters of the beamformer/correlator to account for any
impedance mismatch at the correlator/beamformer input and its gain.
Note that \eqref{eq:S-pass} assumes a 3-port S-parameter model of
hybrids. In the situations when hybrids are 4-port devices requiring
a matched load at the extra port, $S_{\text{H}}$ becomes a 4-port
matrix in \eqref{eq:S-pass}. In this case $S_{\text{PASS}}$ would
also include 1-port S-parameter matrices for each termination.

\subsection{Noise}

For noise analysis, the input signals are removed by setting $\mathbf{a}_{\text{s}}=\left[\mathbf{0}\right]$,
and the noise waves, $\mathbf{b}_{\text{n}}$, emanating from each
port of the network are found from \eqref{eq:b-bs-c} as
\begin{equation}
\left.\mathbf{b}\right|_{\mathbf{a}_{\text{s}}=\left[\mathbf{0}\right]}\equiv\mathbf{b}_{\text{n}}=\left(\mathbf{I}-\mathbf{S}\mathbf{K}\right)^{-1}\mathbf{c}=\mathbf{Q}\mathbf{c},\label{eq:b-vs-n}
\end{equation}
where $\mathbf{Q}$ is introduced for brevity and $\mathbf{I}$ is
the identify matrix. The noise-correlation matrix is described by
\begin{equation}
\overline{\mathbf{b}_{\text{n}}\mathbf{b}_{\text{n}}^{\dagger}}=\mathbf{Q}\overline{\mathbf{c}\mathbf{c}^{\dagger}}\mathbf{Q}^{\dagger}\label{eq:brn}
\end{equation}
where ``$\dagger$'' indicates a Hermitian conjugate operator and,
for the passive components, the terms of $\overline{\mathbf{c}\mathbf{c}^{\dagger}}$
are found from Bosma's theorem \cite{Bosma1967,Wedge1991} as
\begin{equation}
\overline{\mathbf{c}_{\text{PASS}}\mathbf{c}_{\text{PASS}}^{\dagger}}=kB\left(\mathbf{I}-\mathbf{S}_{\text{PASS}}\mathbf{S}_{\text{PASS}}^{\dagger}\right)\mathbf{T}_{\text{PASS}},\label{eq:c-pass}
\end{equation}
where $\mathbf{T}_{\text{PASS}}$ is the diagonal matrix of the physical
temperatures of all passive components. $\mathbf{T}_{\text{PASS}}$
is a $5M\times5M$ matrix for the system in Fig. \ref{fig:Block-diagram}.
Assuming that the LNAs are identical (i.e., noise-correlation matrices
$\overline{\mathbf{c}_{\text{L1}}\mathbf{c}_{\text{L1}}^{\dagger}}=\ldots=\overline{\mathbf{c}_{\text{LM}}\mathbf{c}_{\text{LM}}^{\dagger}}=\mathbf{c}_{\text{L}}$),
the terms in $\overline{\mathbf{c}\mathbf{c}^{\dagger}}$ due to active
components (i.e. $\overline{\mathbf{c}_{\text{ACT}}\mathbf{c}_{\text{ACT}}^{\dagger}}$)
are found from active component noise parameters \cite[p. 54]{Wedge1992}
as
\begin{equation}
\mathbf{c}_{\text{L}}=kT_{\text{L}}B\left[\begin{array}{cc}
\overline{\left|c_{\text{L,}1}\right|^{2}} & \overline{c_{\text{L,}1}c_{\text{L,}2}^{*}}\\
\overline{c_{\text{L,}1}^{*}c_{\text{L,}2}} & \overline{\left|c_{\text{L,}2}\right|^{2}}
\end{array}\right]\label{eq:Rrec-1}
\end{equation}
where \textbf{$T_{\text{L}}$} is the LNA physical temperature\footnote{The assumption here is that the amplifier noise is linearly proportional
to temperature around $T_{0}$= 290 K.} and\textbf{}
\begin{equation}
\left\{ \begin{array}{l}
\overline{\left|c_{\text{L,}1}\right|^{2}}=\frac{T_{min}}{T_{0}}\left(\left|S_{\text{L,}11}\right|^{2}-1\right)+4N\frac{\left|1-S_{\text{L,}11}\Gamma_{opt}\right|^{2}}{1-\left|\Gamma_{opt}\right|^{2}}\\
\overline{\left|c_{\text{L,}2}\right|^{2}}=\left|S_{L,21}\right|^{2}\left(\frac{T_{min}}{T_{0}}+4N\frac{\left|\Gamma_{opt}\right|^{2}}{1-\left|\Gamma_{opt}\right|^{2}}\right)\\
\overline{c_{\text{L,}1}c_{\text{L,}2}^{*}}=\frac{S_{\text{L,}11}}{S_{\text{L,}21}}\overline{\left|c_{\text{L,}2}\right|^{2}}-4N\frac{S_{\text{L,}21}^{*}\Gamma_{opt}^{*}}{1-\left|\Gamma_{opt}\right|^{2}}.
\end{array}\right.\label{eq:def-c}
\end{equation}
The noise parameters in \eqref{eq:def-c} are: the minimum noise temperature,
$T_{\text{min}}$; the Lange invariant, $N$ \cite{Lange1967,Belostotski_MTT2011b};
and the optimum reflection coefficient for minimum noise, $\Gamma_{\text{opt}}$.

Having found $\overline{\mathbf{b}_{\text{n}}\mathbf{b}_{\text{n}}^{\dagger}}$
in \eqref{eq:brn}, we can calculate beam-equivalent receiver noise
temperature, $T_{\text{rec}}$, with \cite{Warnick2010,RoshaanAli_2021}
\begin{equation}
T_{\text{rec}}=T_{0}\frac{\mathbf{\tilde{w}}^{\dagger}\left.\overline{\mathbf{b}_{\text{n}}\mathbf{b}_{\text{n}}^{\dagger}}\right|_{T_{\text{A}}=0}\mathbf{\tilde{w}}}{\tilde{\mathbf{w}}^{\dagger}\left.\overline{\mathbf{b}_{\text{n}}\mathbf{b}_{\text{n}}^{\dagger}}\right|_{\text{only }T_{\text{A}}\neq0}\tilde{\mathbf{w}}},\label{eq:Trec}
\end{equation}
where $\mathbf{\tilde{w}}$ is the vector of complex gains of the
beamformer/correlator in which all terms are zero except for those
pertaining to the elements of $\mathbf{b}$ that relate to power flow
to the beamformer/correlator. The beam-equivalent noise temperature
$T_{\text{rec}}$ is defined with reference to the response of the
antenna arrays to an isotropic external noise environment at temperature
$T_{0}$ with which they are in thermal equilibrium.

Further, we find the cross-correlation between outputs $i$ and $j$
in the form of noise temperature by
\begin{equation}
T_{ij}=\frac{1}{kB}\mathbf{\tilde{w}}_{i}^{\dagger}\left.\overline{\mathbf{b}_{\text{n}}\mathbf{b}_{\text{n}}^{\dagger}}\right|_{\mathbf{a}_{\text{s}}=\left[\mathbf{0}\right]}\mathbf{\tilde{w}}_{j}.\label{eq:Cross-corr}
\end{equation}

\subsection{Gain}

To find gain, we set $\mathbf{c}=\left[\mathbf{0}\right]$ and set
$\mathbf{a}_{\text{s}}$ elements, which correspond to the antenna-array
ports, non-zero to apply source waves at the outputs of array antennas
in \eqref{eq:b-bs-c} to obtain
\begin{equation}
\mathbf{b}=\mathbf{Q}\mathbf{S}\mathbf{a}_{\text{s}}.\label{eq:b-out}
\end{equation}
With $\mathbf{b}$ from \eqref{eq:b-out}, in general the gain of
the system in response to an input is 
\begin{equation}
G=\frac{\mathbf{\tilde{w}}^{\dagger}\overline{\mathbf{b}\mathbf{b}^{\dagger}}\mathbf{\tilde{w}}}{\mathbf{\mathbf{\delta}}^{\dagger}\overline{\mathbf{a}_{\text{s}}\mathbf{a}_{\text{s}}^{\dagger}}\mathbf{\mathbf{\delta}}},\label{eq:Gain}
\end{equation}
where $\delta$ is the vector with non-zero terms pertaining to the
array inputs. This gain definition considers any or all elements of
$\mathbf{a}_{\text{s}}$ as inputs and any or all elements of $\mathbf{b}$
as outputs. In practice, $\mathbf{\delta}$ and $\mathbf{w}$ would
only specify some of their elements, such as those pertaining to antenna
ports as inputs and those pertaining to the waves traveling towards
the beamformer as outputs. Further, the expression of the correlation
gain can be obtained from \eqref{eq:Gain} with 
\begin{equation}
G_{ij}=\frac{\mathbf{\tilde{w}}_{i}^{\dagger}\overline{\mathbf{b}\mathbf{b}^{\dagger}}\mathbf{\tilde{w}}_{j}}{\mathbf{\mathbf{\delta}}_{k}^{\dagger}\overline{\mathbf{a}_{\text{s}}\mathbf{a}_{\text{s}}^{\dagger}}\mathbf{\mathbf{\delta}}_{l}^{\dagger}},\label{eq:Gain-ij}
\end{equation}
where subscripts $k$ and $l=1\ldots M$ indicate elements of input
vector $\mathbf{a}_{\text{s}}$ that form the input to the system,
and subscripts $i$ and $j=1\ldots M$ indicate outputs pertaining
to the correlation gain of interest.

\section{\label{sec:Receiver-noise-temperature:}Two-Element Phased-Array
$T_{\text{rec}}$}

Conventionally, increasing array sensitivity requires the development
of receivers such that their optimum reflection coefficient, $\Gamma_{\text{opt}}\in\mathbb{C}$,
for minimum noise equals the active reflection coefficient, $\Gamma_{\text{act}}\in\mathbb{C}$,
of the array \cite{Maaskant2007,Maaskant2008,Belostotski2009_TAP,Warnick2010,Belostotski_TAP2015}.
For a given receiver, this equality can only be achieved for one scan
angle, which results in only one beam whose noise can be fully minimized.
For other scan angles, $\Gamma_{\text{act}}$ deviates from $\Gamma_{\text{opt}},$
thereby increasing array noise above its minimum. The scan-dependence
of $\Gamma_{\text{act}}$ is due to mutual coupling of antennas in
the array. Because of such coupling, some of the receiver noise couples
to adjacent antennas and flows to the output through the beamformer,
adversely impacting the output noise levels. Past methods of reducing
mutual coupling included decoupling networks and low scattering antennas
\cite{Andersen1976,Wang2004,Hein2006,Kossiavas2006,Hein2008}. The
authors of \cite{Warnick2007} demonstrated that decoupling is necessary
for optimum noise matching of antenna arrays. Predictably, the addition
of the decoupling network introduces noise, thus potentially negating
the advantage of having a decoupled array.

In this section, we investigate the effect of the broadband mutual-coupling
canceler on $T_{\text{rec}}$ of a two-element antenna array and the
possibility of using this canceler for making $\Gamma_{\text{act}}$
scan independent.

\subsection{\label{subsec:Circuit-models}Circuit models}

In the following numerical calculations, we assume LNA $T_{\text{min}}=25\,\text{K}$,
$N=0.03$, and $\Gamma_{\text{opt}}=0.2\angle100^{\circ}$. Because
both $T_{\text{min}}$ and $N$ are invariant upon lossless embedding,
$\Gamma_{\text{opt}}$ can be varied with a matching network without
affecting $T_{\text{min}}$ and $N$. We will use this fact to search
for the optimum $\Gamma_{\text{opt}}$ that minimizes $T_{\text{rec}}$.
The S-parameters of the LNA are assumed as
\begin{equation}
\mathbf{S}_{\text{L}}=\left[\begin{array}{cc}
0.2\angle-75^{\circ} & 0.01\angle150^{\circ}\\
3\angle-150^{\circ} & 0.3\angle-100^{\circ}
\end{array}\right].\label{eq:SL}
\end{equation}
We based the antenna array S-parameters on a scaled version of a
30-mm spaced two-element array described in \cite{deSilva2020}:
\begin{equation}
\mathbf{S}_{\text{A}}=\mathbf{S}_{\text{R}}=\left[\begin{array}{cc}
0.3\angle100^{\circ} & 0.2\angle-60^{\circ}\\
0.2\angle-60^{\circ} & 0.3\angle100^{\circ}
\end{array}\right].
\end{equation}
The ideal hybrid S-parameters are
\begin{equation}
\mathbf{S}_{\text{H}}=\frac{1}{\sqrt{2}}\left[\begin{array}{ccc}
0 & 1\angle P_{\text{H}} & 1\\
1\angle P_{\text{H}} & 0 & 0\\
1 & 0 & 0
\end{array}\right],
\end{equation}
where $P_{\text{H}}=90^{\circ}$, port 1 is the common port and ports
2 and 3 are the $90^{\circ}$ and $0^{\circ}$ coupled ports. For
calculations with a non-ideal hybrid we use the manufacturer-specified
S-parameters of a commercial component (Mini-Circuits JSPQW-100A+)
at 100 MHz.

\subsection{Numerical results}

\begin{figure}
\subfloat[]{\includegraphics[width=0.5\columnwidth]{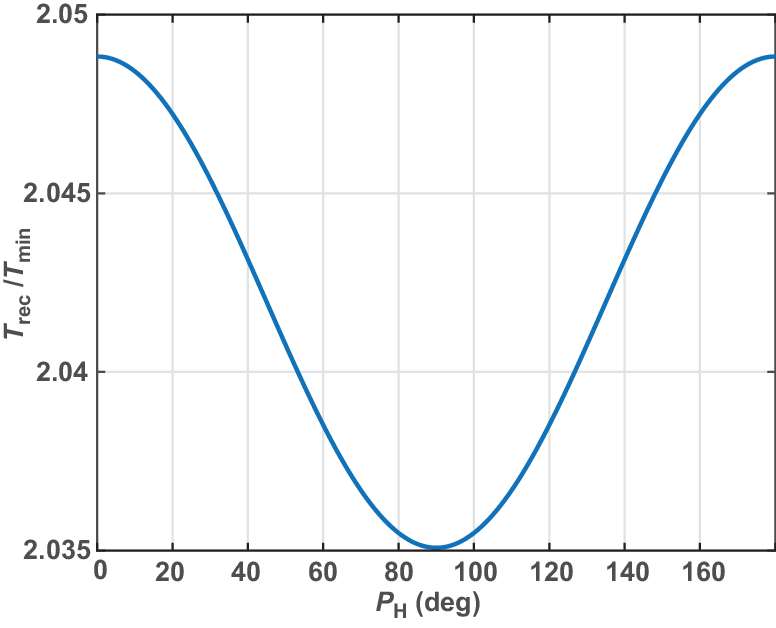}

}\subfloat[]{\includegraphics[width=0.5\columnwidth]{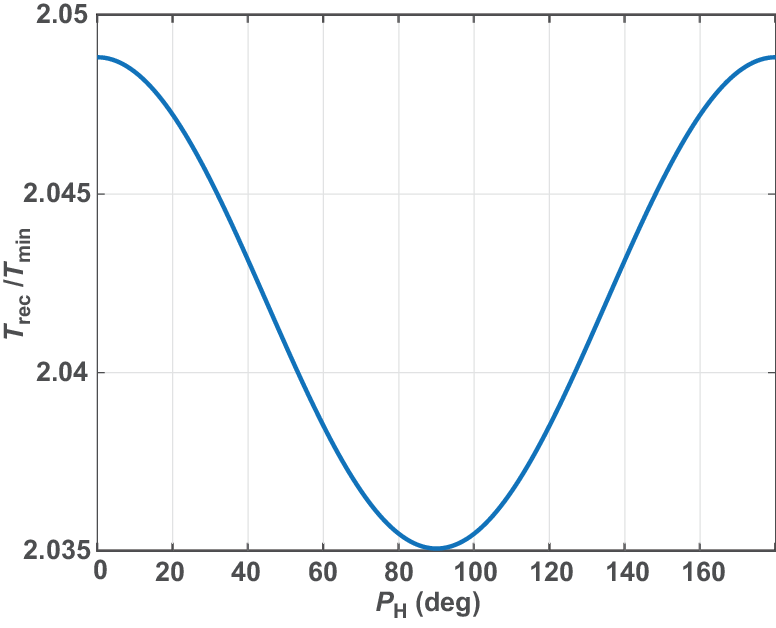}

}

\subfloat[]{\includegraphics[width=0.5\columnwidth]{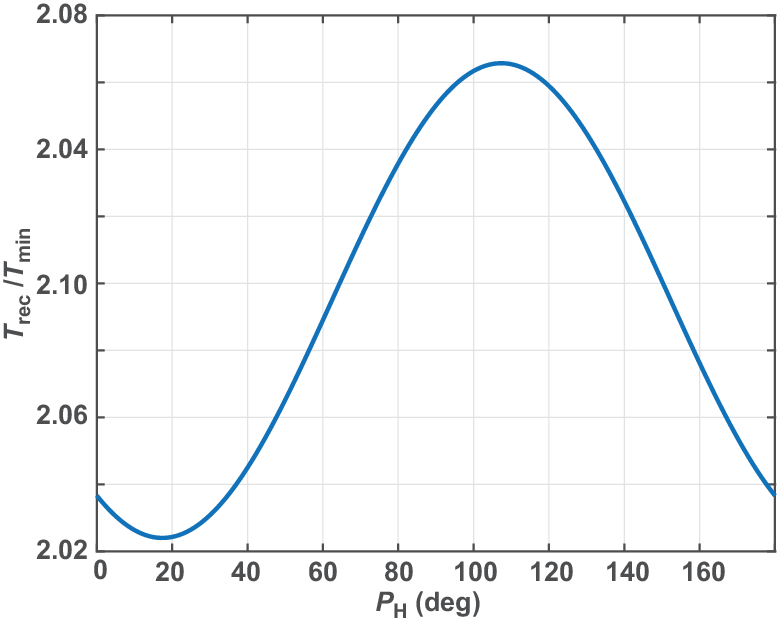}

}\subfloat[]{\includegraphics[width=0.5\columnwidth]{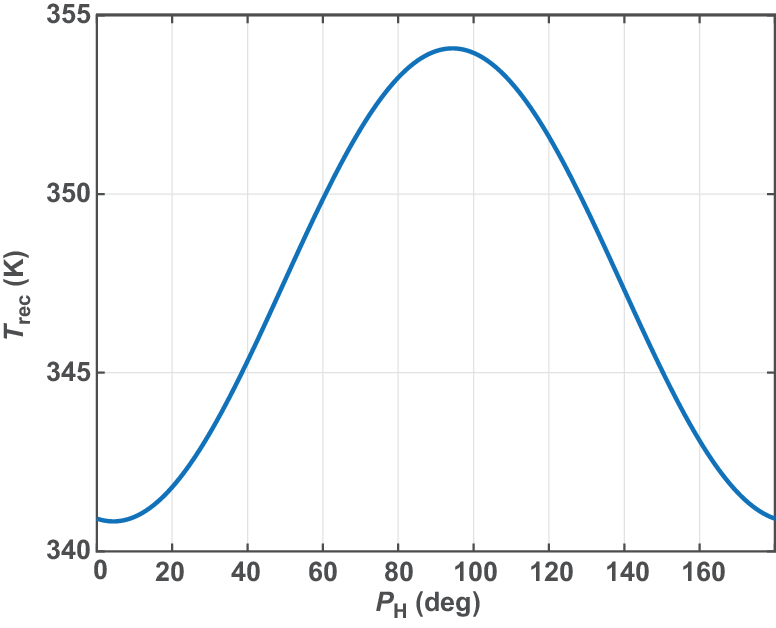}

}

\caption{\label{fig:Sim-results-LNA-coupling}$T_{\text{rec}}$ for the system
configured as: a) $\Gamma_{\text{opt}}=0$, $S_{\text{L},11}=0$,
ideal hybrids, and $T_{\text{H}}=T_{\text{R}}=0\,\text{K}$; b) Same
as a) but with $S_{\text{L},11}=0.2\angle-75^{\circ}$; c) Same as
b) but $\Gamma_{\text{opt}}=0.2\angle-100^{\circ}$;  and d) same
as c) but with all temperatures in $\mathbf{T}_{\text{PASS}}$ set
to $290\,\text{K}$. Since in subfigure d) $T_{\text{rec}}$ is dominated
by the ambient temperature rather than $T_{\text{min}}$, the y-axis
is not normalized.}
\end{figure}

We first start with an ideal model in which we set LNA $\Gamma_{\text{opt}}=0$
and $S_{L,11}=0$, use ideal hybrids, and sweep their phase $P_{\text{H}}$
from $0^{\circ}$ to $180^{\circ}$ while keeping the physical temperatures
of the antenna array and the LNA at $290\,\text{K}$ and all other
temperatures in $\mathbf{T}_{\text{PASS}}$ at zero. The outcome of
this baseline calculation is shown in Fig. \ref{fig:Sim-results-LNA-coupling}(a).
As can be seen, $P_{\text{H}}=90^{\circ}$ does indeed reduce $T_{\text{rec}}$
as expected because mutual coupling between antennas is canceled.
We also observe that the minimum $T_{\text{rec}}$ is slightly higher
than $2T_{\text{min}}$. This is due to hybrid loss, $L_{\text{H}}$,
and the impedance mismatch between the antenna and the hybrid. As
a result, the variation of $T_{\text{rec}}$ with $P_{\text{H}}$
is very insignificant. When $S_{\text{L},11}$ is set to its value
in \eqref{eq:SL}, the resultant $T_{\text{rec}}$ remains unchanged
as shown in Fig. \ref{fig:Sim-results-LNA-coupling}(b).

Next, we assigned $\Gamma_{\text{opt}}=0.2\angle-100^{\circ}$ while
keeping everything else unchanged. Fig. \ref{fig:Sim-results-LNA-coupling}(c)
shows that $P_{\text{H}}=90^{\circ}$ is no longer optimum. Instead,
$P_{\text{H}}\approx17.5^{\circ}$ now realizes low $T_{\text{rec}}$.
At $P_{\text{H}}\approx17.5^{\circ}$, mutual coupling is not canceled
and beamformer dependent $T_{\text{rec}}$ results. However, $T_{\text{rec}}$
is reduced slightly from that in Figs. \ref{fig:Sim-results-LNA-coupling}(a)
and (b) demonstrating that the removal of mutual coupling is not necessarily
required for improving the overall noise of the receiver under investigation.

Fig. \ref{fig:Sim-results-LNA-coupling}(d) plots $T_{\text{rec}}$
when temperatures of all components are set to $290\,\text{K}$. In
this case, the replica array with the hybrids contributes $290\,\text{K}$
in addition to $\sim L_{\text{H}}\times T_{\text{min}}$ from the
LNA. As shown in these calculations, $T_{\text{rec}}$ is significantly
higher than $T_{\text{min}}$, and, therefore, from the array sensitivity
point of view, this type of canceler network does not provide the
benefits of high sensitivity. Based on this analysis and the general
expectation that the replica array contributes $T_{\text{R}}$ to
$T_{\text{rec}}$, $T_{\text{rec}}$ cannot be less than $T_{\text{R}}$
even if the hybrids are designed such that $L_{\text{H}}=0\,\text{dB}$.
Comparing Figs. \ref{fig:Sim-results-LNA-coupling}(c) and (d) suggests
that in order to maintain $T_{\text{rec}}$ between $T_{\text{min}}$
and $L_{\text{H}}T_{\text{min}}$, the passives should be cooled to
temperature much lower than $290\,\text{K}$. Reducing $L_{\text{H}}$
would further help reducing $T_{\text{rec}}$.

\begin{figure}
\subfloat[]{\includegraphics[bb=0bp 0bp 257bp 257bp,clip,width=0.5\columnwidth]{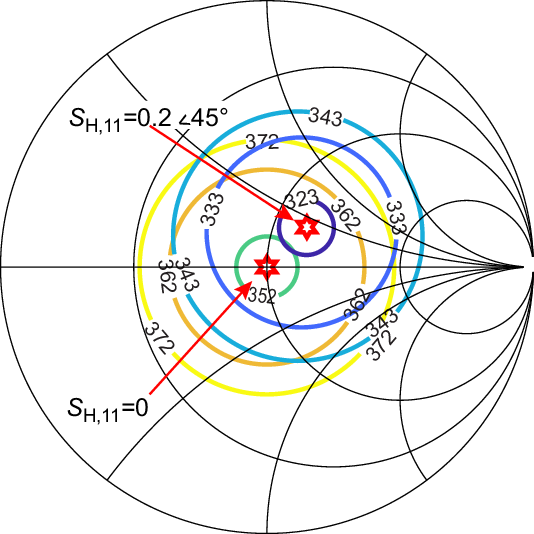}

}\subfloat[]{\includegraphics[bb=0bp 0bp 258bp 258bp,clip,width=0.5\columnwidth]{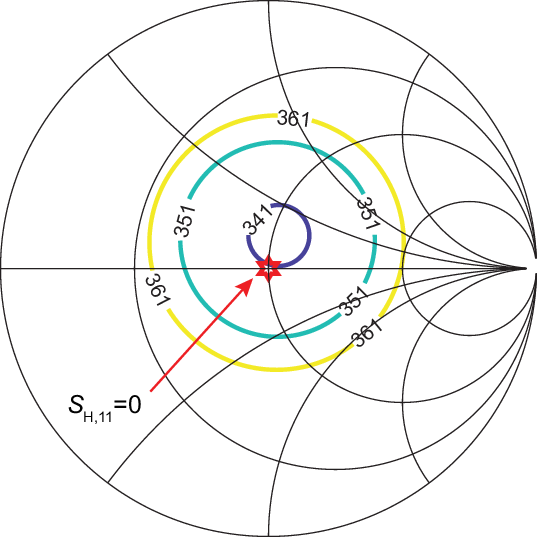}

}

\caption{\label{fig:Optimum-LNA}Contours of constant $T_{\text{rec}}$ as
functions of $\Gamma_{\text{opt}}$ at: a) $P_{\text{H}}=90^{\circ}$
for $S_{\text{H},11}=0$ and $0.2\angle45^{\circ}$; and b) $P_{\text{H}}=4.4^{\circ}$
from \ref{fig:Sim-results-LNA-coupling}(d) and for $S_{\text{H},11}=0$.}
\end{figure}

In subsequent calculations, we set $P_{\text{H}}=90^{\circ}$ to investigate
the independence of optimum $\Gamma_{\text{opt}}$ on the values of
beamformer coefficients. As shown by the contours of constant noise
temperature in Fig. \ref{fig:Optimum-LNA}(a), the value of 
 $\Gamma_{\text{opt}}$ that exhibits the lowest $T_{\text{rec}}$
coincides with $S_{\text{H},11}$ for two different values of $S_{\text{H},11}$:
$S_{\text{H},11}=0$ and $S_{\text{H},11}=0.2\angle45^{\circ}$. Similar
calculations show that the lowest $T_{\text{rec}}$ always occurs
when $\Gamma_{\text{opt}}=S_{\text{H},11}$ for any scan angle, i.e.
$\Gamma_{\text{act}}=S_{\text{H},11}$. It is important to note that
while $\Gamma_{\text{act}}$ is scan-angle independent, $T_{\text{rec}}$
does depend on the scan angle when the reflection coefficients of
antennas are not zero.  When $P_{\text{H}}=4.4^{\circ}$ is selected
from Fig. \ref{fig:Sim-results-LNA-coupling}(d) to minimize $T_{\text{rec}}$,
the optimum $\Gamma_{\text{opt}}\neq S_{\text{H},11}$, as seen from
Fig. \ref{fig:Optimum-LNA}(b), and the array is no longer decoupled.

\begin{figure}
\begin{centering}
\includegraphics[bb=0bp 0bp 258bp 258bp,clip,width=0.5\columnwidth]{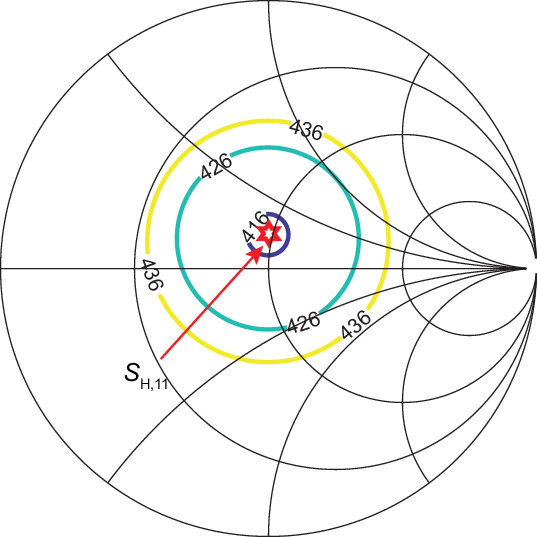}
\par\end{centering}
\caption{\label{fig:Trec-nothing-ideal}Contours of constant $T_{\text{rec}}$
as function of $\Gamma_{\text{opt}}$ of the network with the non-ideal
hybrid.}
\end{figure}

In the last step of this section, we replace the S-parameters of the
ideal hybrid with the 100-MHz S-parameters of the commercial hybrid
(Mini-Circuits JSPQW-100A+) . The outcome of this calculation in Fig.
\ref{fig:Trec-nothing-ideal}
 shows the minimum of $T_{\text{rec}}$ higher than in Fig. \ref{fig:Optimum-LNA}
due to the non-ideal hybrid while the network remains decoupled, and
$\Gamma_{\text{opt}}=S_{\text{H},11}$ still minimizes $T_{\text{rec}}$.

\section{\label{sec:Output-cross-correlation}Two-Element Mutual Coherence}

\subsection{\label{subsec:Single-Frequency-Two-Element-Int}Single-Frequency
Two-Element Interferometer}

\subsubsection{\label{subsec:Ideal-Hybrids}Ideal Hybrids}

\begin{figure}
\subfloat[]{\includegraphics[width=0.5\columnwidth]{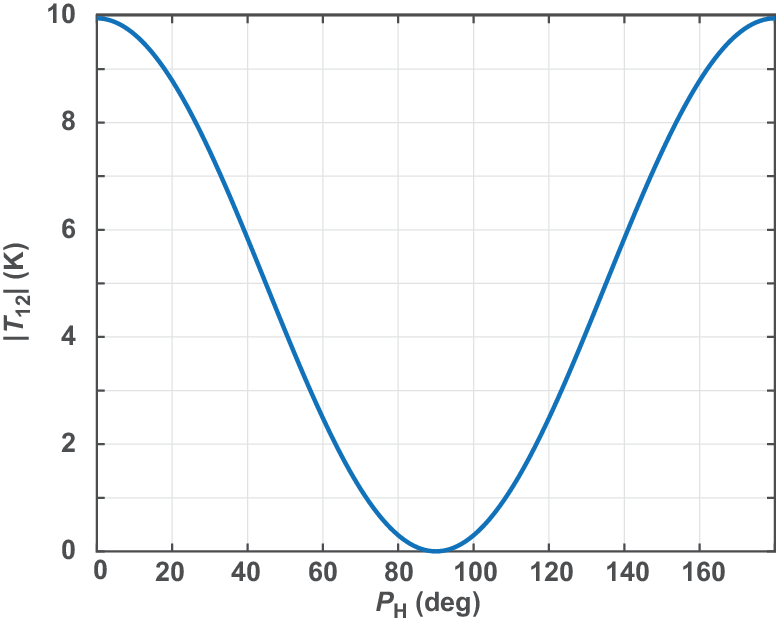}

}\subfloat[]{\includegraphics[width=0.5\columnwidth]{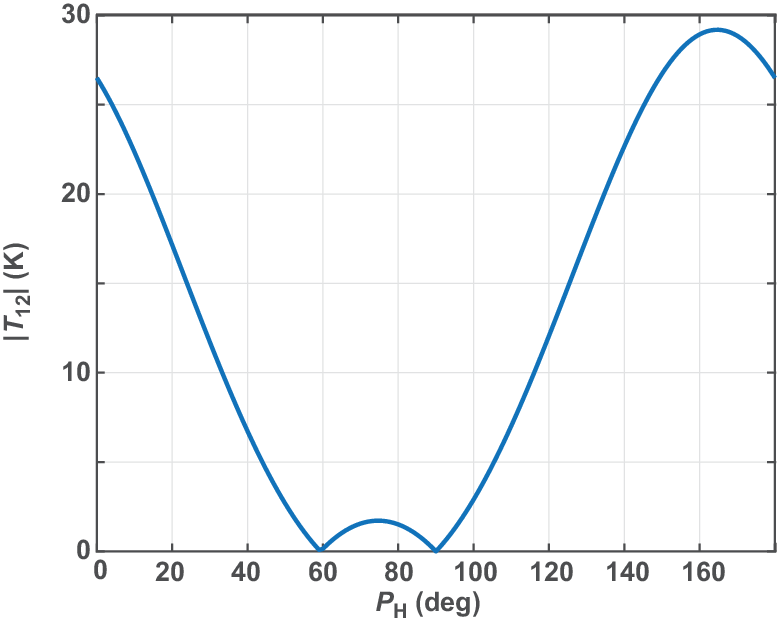}

}

\subfloat[]{\includegraphics[width=0.5\columnwidth]{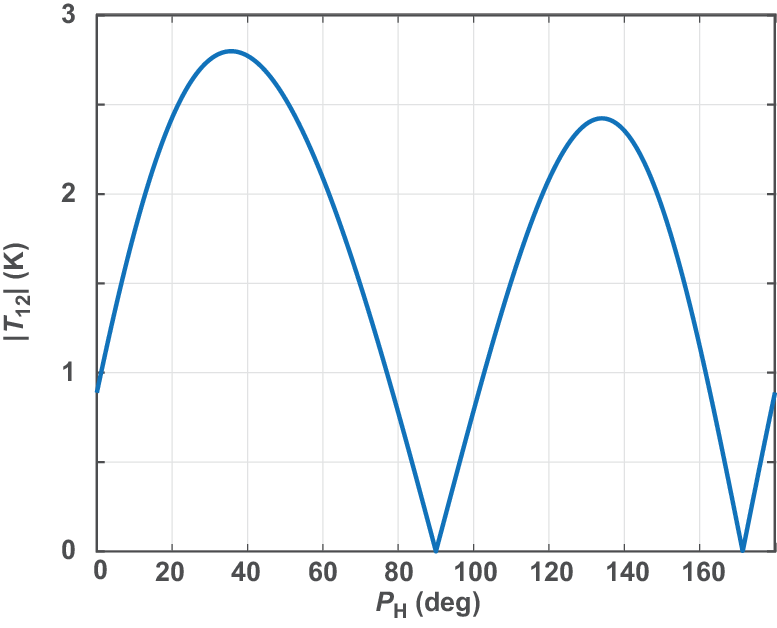}

}\subfloat[]{\includegraphics[width=0.5\columnwidth]{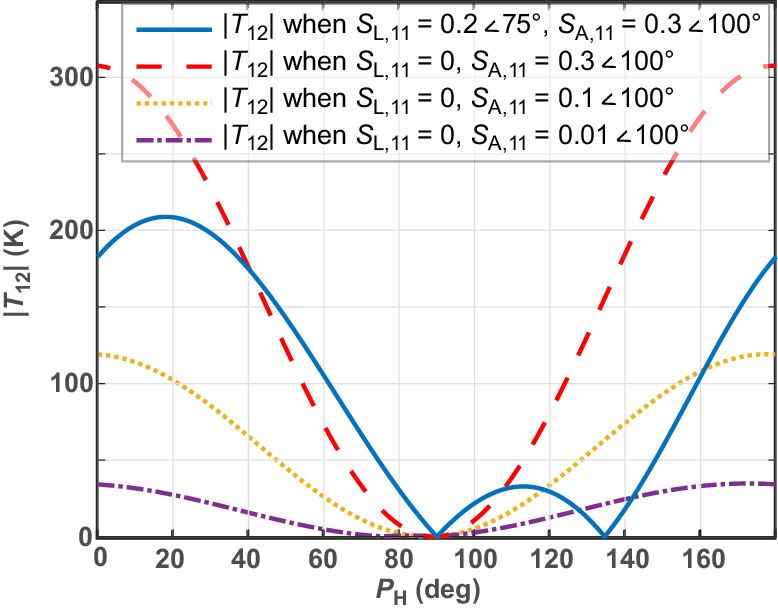}

}

\caption{\label{fig:Sim-results-cross-corr}$\left|T_{\text{12}}\right|$ for
the system configured as: a) $\Gamma_{\text{opt}}=0$, $S_{\text{L},11}=0$,
ideal hybrids, and $T_{\text{A}}=T_{\text{H}}=T_{\text{R}}=0\,\text{K}$;
b) Same as a) but with $S_{\text{L},11}=0.2\angle-75^{\circ}$; c)
Same as b) but $\Gamma_{\text{opt}}=0.2\angle-100^{\circ}$;  and
d) same as c) but with all temperatures in $\mathbf{T}_{\text{PASS}}$
set to $290\,\text{K}$ with two different setting for $S_{\text{L},11}$.}
\end{figure}

In this section, we reuse the same calculation setup as in Fig. \ref{fig:Sim-results-LNA-coupling}
and determine the mutual coherence, in terms of $\left|T_{12}\right|$,
of the two-element array with circuit components as in Section \ref{subsec:Circuit-models}.
Fig. \ref{fig:Sim-results-cross-corr} shows the results of this numerical
analysis.

As can be seen from the sub-figures, it is possible to adjust the
phase of hybrids to reduce cross correlation to zero. As expected,
 $P_{\text{H}}=90^{\circ}$ is optimum, see Fig. \ref{fig:Sim-results-cross-corr}.
However, Fig. \ref{fig:Sim-results-cross-corr}(b) shows that due
to non-zero $S_{\text{L},11}$ a null in the cross-correlation also
exists at $P_{\text{H}}\approx60^{\circ}$. The appearance of the
second null can be explained with the following gedankenexperiment.
Each LNA generates noise at both inputs, $I_{1}$ and $I_{2}$, to
the cross correlator. The noise contributions of LNA\#1 and LNA\#2
to $I_{1}$ are denoted by uncorrelated $s_{1,1}$ and $s_{2,1}$,
respectively, and their contributions to $I_{2}$ are denoted by uncorrelated
$s_{1,2}$ and $s_{2,2}$. The output of the correlator is $\overline{I_{1}^{*}I_{2}}=\overline{\left(s_{1,1}+s_{2,1}\right)^{*}\left(s_{1,2}+s_{2,2}\right)}$,
which is equivalent to $\overline{I_{1}^{*}I_{2}}=\overline{s_{1,1}^{*}s_{1,2}}+\overline{s_{1,1}^{*}s_{2,2}}+\overline{s_{2,1}^{*}s_{1,2}}+\overline{s_{2,1}^{*}s_{2,2}}$,
where the first term, pertaining to the noise of LNA\#1, and the last
terms, pertaining to LNA\#2, are not zero. The other two terms are
cross-correlations of LNA\#1 and LNA\#2 generated noise and therefore
are zero. The locations of the two LNAs relative to $I_{1}$ and $I_{2}$
are different. It is expected that since for identical LNAs $\overline{s_{1,1}^{*}s_{1,2}}=\overline{s_{2,1}s_{2,2}^{*}}$
, then $\overline{I_{1}^{*}I_{2}}=2\Re\left\{ \overline{s_{1,1}^{*}s_{1,2}}\right\} $,
and it is possible to null the cross correlation by arranging $P_{\text{H}}$
such that $\overline{s_{1,1}^{*}s_{1,2}}$ is purely imaginary. At
$P_{\text{H}}\approx60^{\circ}$, $\overline{s_{1,1}^{*}s_{1,2}}$
is in fact purely imaginary causing low cross correlation. While
the decoupling at $P_{\text{H}}=90^{\circ}$ due to hybrids is unique
to the described system, the extra null in mutual coherence at $P_{\text{H}}\approx60^{\circ}$
is not unique and, for example, is present in (11) from \cite{SutinjoURI2020}.

$\Gamma_{\text{opt}}$ is also responsible for moving the location
of the additional null as can be seen by comparing Figs. \ref{fig:Sim-results-cross-corr}(b)
and (c). Further, having other system components contribute noise
modifies the location of the extra null. The resultant $P_{\text{H}}$
is dependent on $S_{\text{L},11}$ and $S_{\text{A},11(22)}$ as demonstrated
in Fig. \ref{fig:Sim-results-cross-corr}(d). As $S_{\text{A},11(22)}$
approaches zero, the maximum of $\left|T_{12}\right|$ reduces because
noise waves emanating from matched array ports connected to ideal
hybrids exhibit no cross-correlation. However, the extra null still
exists in this case albeit it is not very visible in the figure. Moreover,
the array mutual coupling affects $\left|T_{12}\right|$. As the amount
of coupling increases the peaks of the cross correlation function
increase. Finally, these simulations show that a well-matched LNA
with $\left|S_{\text{L},11}\right|\approx0$ avoids sharp nulls in
$\left|T_{12}\right|$ and creates a smooth trough instead thereby
reducing the sensitivity of $\left|T_{12}\right|$ to $P_{\text{H}}$.

\begin{figure}
\begin{centering}
\includegraphics[width=0.7\columnwidth]{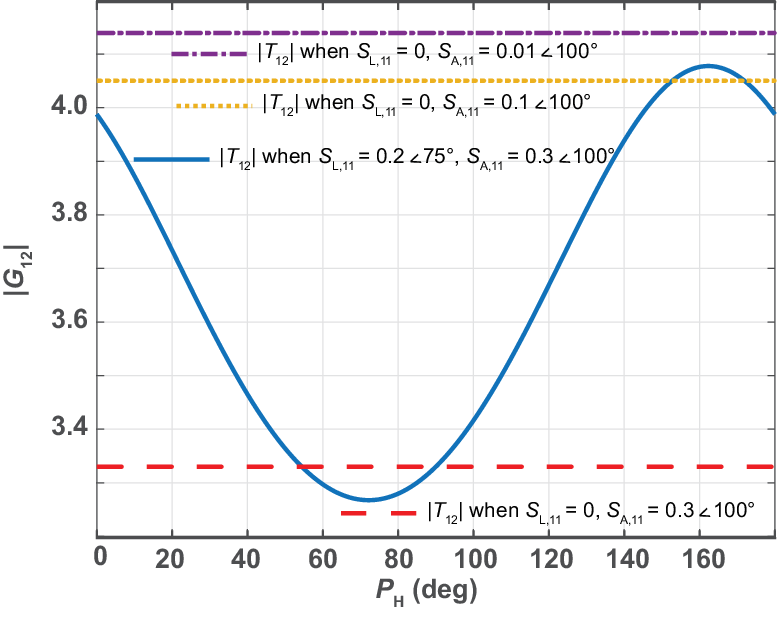}
\par\end{centering}
\caption{\label{fig:Gain-as-function-phase}Magnitude of correlation gain,
$\left|G_{12}\right|$, (linear units) as a function of $P_{\text{H}}$
for different $S_{\text{L},11}$ and $S_{\text{A},11(22)}$.}
\end{figure}

Based on the same model settings as in Fig. \ref{fig:Sim-results-LNA-coupling}(d),
Fig. \ref{fig:Gain-as-function-phase} also confirms that the correlation
gain, $G_{12}$, of the network to a correlated input signal is not
zero even at $P_{\text{H}}$ where receiver-network-noise cross-correlation
is nulled. 
 To generate the input signal for this analysis, we make use of the
results in \cite{Warnick2008}, which showed that noise correlation
matrix of an array due to thermal radiation from the warm scene can
be found via Bosma's theorem. Therefore, to find $G_{12}$, the array
is excited with an input $\mathbf{a}_{\text{s}}$ whose noise-correlation
matrix $\overline{\mathbf{a}_{\text{s}}\mathbf{a}_{\text{s}}^{\dagger}}=\left[\begin{array}{cc}
\overline{\mathbf{c}_{\text{PASS}}\mathbf{c}_{\text{PASS}}^{\dagger}} & \overline{\mathbf{c}_{\text{L}}\mathbf{c}_{\text{L}}^{\dagger}}\end{array}\right]$. We set $T_{\text{R}}=T_{\text{H}}=T_{\text{L}}=0$. By doing so,
only the antenna array excites the network, and other network components
are prevented from adding to the output power of the interferometer
and inadvertently increasing the computed gain. As the result $\overline{\mathbf{a}_{\text{s}}\mathbf{a}_{\text{s}}^{\dagger}}=\left[\begin{array}{cc}
\overline{\mathbf{c}_{\text{A}}\mathbf{c}_{\text{A}}^{\dagger}} & \mathbf{0}\end{array}\right]$, and the correlation of the input signal received by the array is
captured through the cross-correlation terms of $\overline{\mathbf{a}_{\text{s}}\mathbf{a}_{\text{s}}^{\dagger}}$.
While there is some phase dependence due to $S_{\text{L},11}$, the
gain $G_{12}$, as found from \eqref{eq:Gain-ij}, remains non-zero
and approximately near $\left|S_{\text{L},21}\right|^{2}/L_{\text{H}}$
for all $P_{\text{H}}$. As $\left|S_{\text{A},11(22)}\right|$ decreases
the hybrid insertion loss decreases towards its minimum value of 2,
thereby increasing $G_{12}$ until its maximum value.

While the correlated noise from the replica array is canceled, $G_{12}\neq0$
for the correlated output from the antenna array. To explain this,
we consider a situation in which both arrays are placed in separate
absorbing chambers at thermal equilibrium with all other passive components
such that $T_{\text{A}}=T_{\text{R}}=T_{\text{H}}$. In this case,
the noises at the hybrid outputs are uncorrelated. If we now remove
the absorbing chamber surrounding the antenna array, $\overline{\mathbf{c}_{\text{A}}\mathbf{c}_{\text{A}}^{\dagger}}$
changes from $\overline{\mathbf{c}_{\text{A}}\mathbf{c}_{\text{A}}^{\dagger}}=kB\left(\mathbf{I}-\mathbf{S}_{\text{A}}\mathbf{S}_{\text{A}}^{\dagger}\right)\mathbf{T}_{\text{A}}$
to $\left[kB\left(\mathbf{I}-\mathbf{S}_{\text{A}}\mathbf{S}_{\text{A}}^{\dagger}\right)\mathbf{T}_{\text{A}}-\mathbf{C}_{\text{ext}}\left(T_{\text{A}}\right)\right]+\mathbf{C}_{\text{ext}}\left(T_{\text{sky}}\right)$,
where $\mathbf{C}_{\text{ext}}$ is the contribution from the external
isotropic thermal noise and the first two terms form the noise contribution
due to ohmic losses of the antenna array \cite{Warnick2008}. Consequently,
 $\overline{\mathbf{b}\mathbf{b}^{\dagger}}$ leading to \eqref{eq:Gain-ij}
is written as
\begin{align}
\left.\overline{\mathbf{b}\mathbf{b}^{\dagger}}\right|_{T_{\text{L}}=0} & =\mathbf{Q}\mathbf{S}\overline{\mathbf{a}_{\text{s}}\mathbf{a}_{\text{s}}^{\dagger}}\mathbf{S}^{\dagger}\mathbf{Q}^{\dagger}\\
 & =\mathbf{Q}\mathbf{S}\left[\begin{array}{cc}
\mathbf{C}_{\text{ext}}\left(T_{\text{sky}}\right)-\mathbf{C}_{\text{ext}}\left(T_{\text{A}}\right) & \mathbf{0}_{2M\times5M}\\
\mathbf{0}_{5M\times2M} & \mathbf{0}_{5M\times5M}
\end{array}\right]\mathbf{S}^{\dagger}\mathbf{Q}^{\dagger}\nonumber 
\end{align}
to show that, when $T_{\text{sky}}\neq T_{\text{A}}$, the output
of the networks is non-zero, resulting in a non-zero gain. In conclusion,
the accuracy of detecting $T_{\text{sky}}$ therefore relies on the
accuracy of measuring $T_{\text{A}}$.

\subsubsection{\label{subsec:Non-Ideal-Hybrids}Non-Ideal Hybrids}

\begin{figure}
\subfloat[]{\includegraphics[width=0.5\columnwidth]{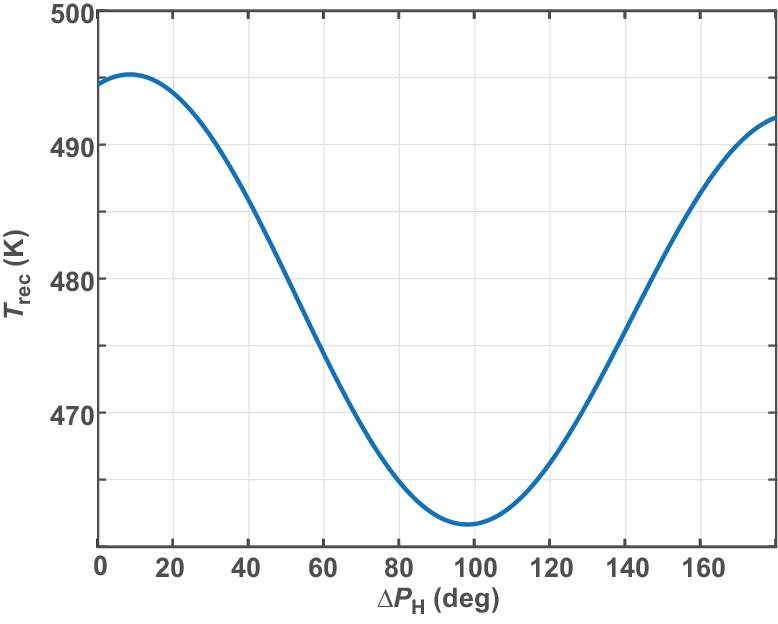}

}\subfloat[]{\includegraphics[width=0.5\columnwidth]{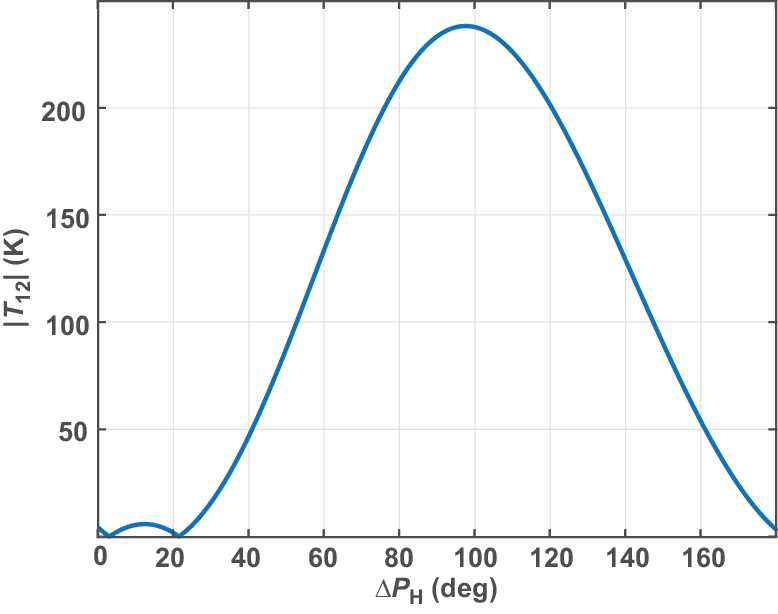}

}

\subfloat[]{\includegraphics[bb=0bp 0bp 258bp 258bp,clip,width=0.5\columnwidth]{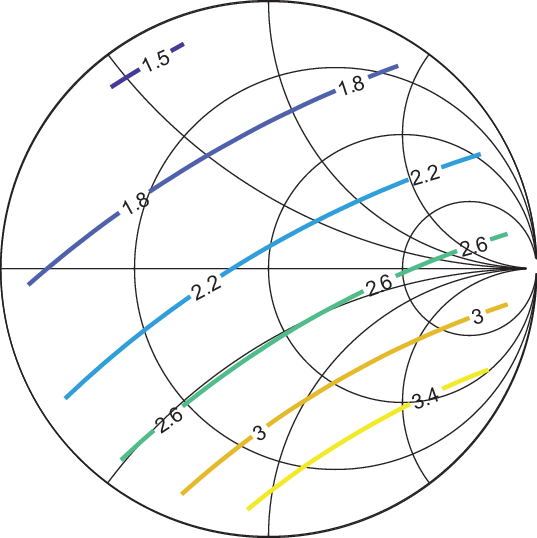}

}\subfloat[]{\includegraphics[bb=0bp 0bp 258bp 258bp,clip,width=0.5\columnwidth]{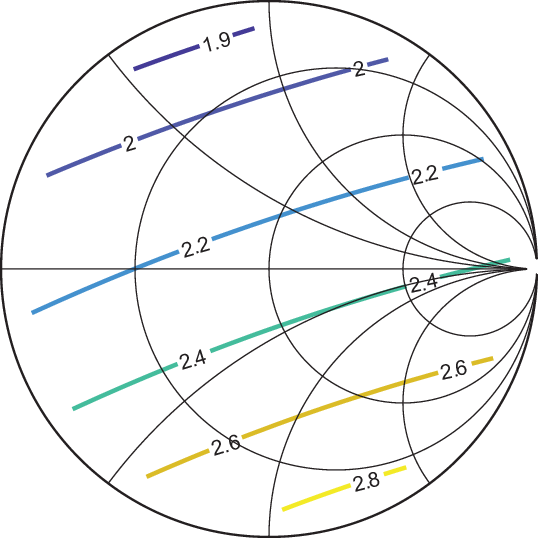}

}

\subfloat[]{\includegraphics[bb=0bp 0bp 258bp 258bp,clip,width=0.5\columnwidth]{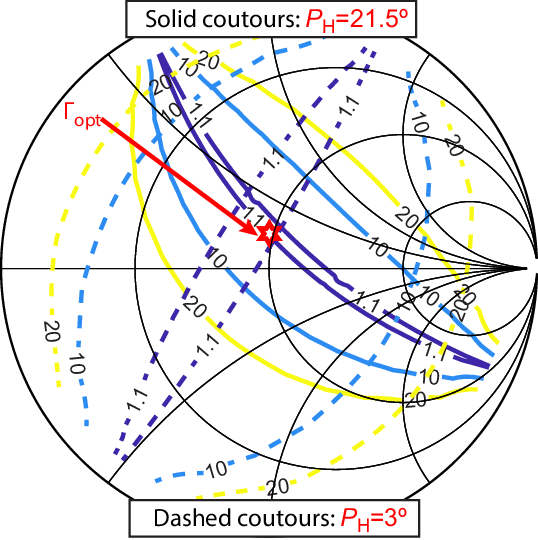}

}\subfloat[]{\includegraphics[bb=0bp 0bp 258bp 258bp,clip,width=0.5\columnwidth]{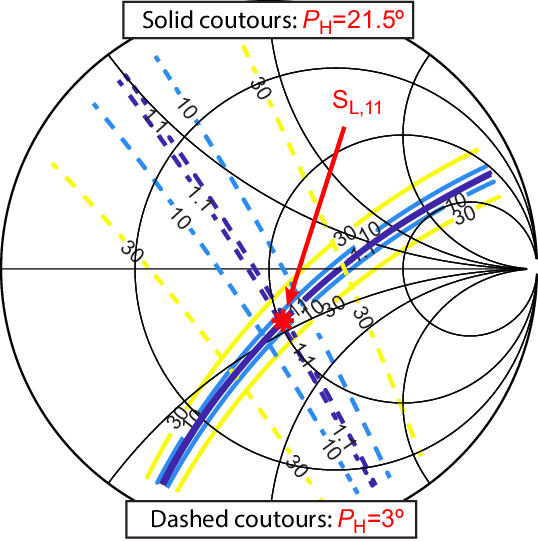}

}

\caption{\label{fig:CrossCorr-nothing-ideal}Simulated a) $T_{\text{rec}}$
and b) $\left|T_{12}\right|$ as functions of $\Delta P_{\text{H}}$
with a non-ideal hybrid. The contour of constant gain $\left|G_{12}\right|$
as functions of $S_{\text{L},11}$ at $\Delta P_{\text{H}}\approx21.5^{\circ}$and
$P_{\text{H}}\approx3^{\circ}$ are shown in c) and d), respectively.
Also shown the contours of constant $\left|T_{12}\right|$ at $\Delta P_{\text{H}}\approx21.5^{\circ}$and
$3^{\circ}$ as functions of e) $\Gamma_{\text{opt}}$ and f) $S_{\text{L},11}$.}
\end{figure}

Next, we used the 100-MHz S-parameters for Mini-Circuits JSPQW-100A+
hybrid in calculations depicted in Fig. \ref{fig:CrossCorr-nothing-ideal}.
As we observed above, the optimum $P_{\text{H}}$ depends on system
components. Therefore, it is expected that in the final system a means
of fine tuning $P_{\text{H}}$ is required. In the calculation results
depicted in Fig. \ref{fig:CrossCorr-nothing-ideal}(a) and (b), we
include two identical phase shifters for each hybrid with adjustable
phases $\Delta P_{\text{H},1}=\Delta P_{\text{H},2}=\Delta P_{\text{H}}$
to determine that for the given system $\Delta P_{\text{H}}\approx98^{\circ}$
and $\Delta P_{\text{H}}\approx3^{\circ}$or $21.5^{\circ}$ are required
to minimize $T_{\text{rec}}$ and $\left|T_{12}\right|$, respectively.
 Figs. \ref{fig:CrossCorr-nothing-ideal}(c) and (d) show the contours
of constant gain magnitude, $\left|G_{12}\right|$, in response to
a correlated input as functions of $S_{\text{L},11}$ at $\Delta P_{\text{H}}\approx21.5^{\circ}$
and $3^{\circ}$. The maximum $\left|G_{12}\right|$ of $\sim3$ is
near the edges of the Smith chart with $\left|G_{12}\right|>0$ throughout
the Smith chart. $\left|G_{12}\right|$ can be increased further by
increasing LNA $S_{\text{L},21}$. From these figures, it is concluded
that the network does not null the correlated input signal while nulling
noise generated in the network circuit components. Note that not all
values of $S_{\text{L},11}$ in Figs. \ref{fig:CrossCorr-nothing-ideal}(c)
and (d) realize nulls in $\left|T_{12}\right|$, and, therefore, not
all $S_{\text{L},11}$ identified in these figures may be desirable.

The contour plots of constant $\left|T_{12}\right|$ as functions
of $\Gamma_{\text{opt}}$ and $S_{\text{L},11}$ are presented in
Figs. \ref{fig:CrossCorr-nothing-ideal}(e) and (f) at $\Delta P_{\text{H}}\approx21.5^{\circ}$
and $3^{\circ}$. The two different phase settings result in two
distinct contour plots.  These calculations show that the phases,
rather than magnitudes, of $\Gamma_{\text{opt}}$ and $S_{\text{L},11}$
are more significant to minimizing $\left|T_{12}\right|$. It is also
observed that the decoupling condition of $\Delta P_{\text{H}}\approx3^{\circ}$
is less sensitive to $\Gamma_{\text{opt}}$ and $S_{\text{L},11}$
than the noise canceling $\Delta P_{\text{H}}\approx21.5^{\circ}$.

\begin{figure}
\subfloat[]{\includegraphics[width=0.5\columnwidth]{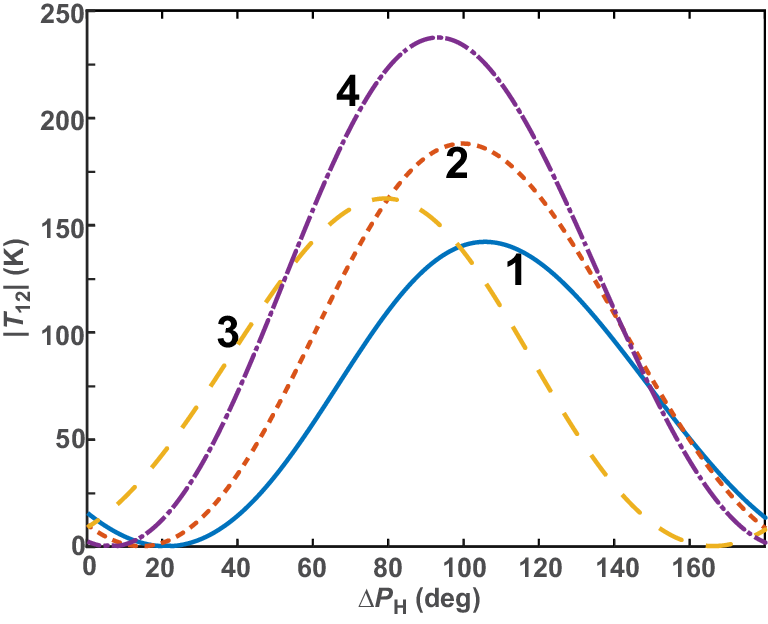}

}\subfloat[]{\includegraphics[bb=0bp 0bp 258bp 258bp,clip,width=0.5\columnwidth]{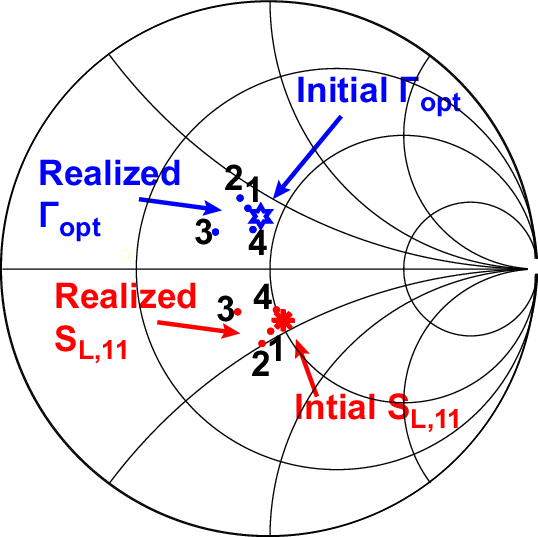}

}

\caption{\label{fig:Wide-Tij-notch}a) Simulated $\left|T_{12}\right|$ with
nearly coincidental zeros ($<4^{\circ}$ difference) in mutual coherence
for lower sensitivity to $\Delta P_{\text{H}}$. b) Locations 1 to
4 of realized $\Gamma_{\text{opt}}$s and $S_{\text{L},11}$s corresponding
to curves identified in a) with corresponding numbers. Also the initial
$\Gamma_{\text{opt}}$ and $S_{\text{L},11}$ of the LNAs are shown.}
\end{figure}

The possibility of adding a matching network to realize coincidental
$\Delta P_{\text{H}}$ for the two nulls in the mutual coherence is
explored in the next experiment whereby identical lossless single-stub
matching networks are placed between the hybrids and LNAs. The matching
network consists of a transmission line with a maximum length of $1\lambda$
that is manipulated in $0.1\lambda$ steps and a shunt capacitor stepped
by 10~pF from 1\,pF to 1\,nF. Four results of this analysis, when
the two nulls in the mutual coherence are located within $4^{\circ}$
of each other, are shown in Fig. \ref{fig:Wide-Tij-notch}
where the resultant $T_{12}$ as well as the realized $\Gamma_{\text{opt}}$
and $S_{\text{L},11}$ are presented.

\subsubsection{Monte-Carlo Sensitivity Analysis}

Sections \ref{subsec:Ideal-Hybrids} and \ref{subsec:Non-Ideal-Hybrids}
progressed from ideal networks to networks that included practical
component models. When non-ideal hybrids were used, two identical
phase shifters, with phases $\Delta P_{\text{H},1}=\Delta P_{\text{H},2}=\Delta P_{\text{H}}$,
were employed to adjust hybrid phases to realize nulls in mutual coherence.
Section \ref{subsec:Non-Ideal-Hybrids} also investigated the possibility
of using a matching network ahead of LNAs so as to make the two nulls
nearly coincidental for relaxed sensitivity to hybrid phase errors.
While these sections assumed that the arrays of antennas, hybrids,
and LNAs were perfectly identical, in practical systems, this assumption
will be violated. For example, imperfect replica-array absorber, fabrication
inaccuracies, process variations in the LNA circuit components, and
so on will cause the loss of identicalness and will require some degree
of tuning and possibly sorting of parts to identify the most similar
ones.  This section analyzes the impact on $\left|T_{12}\right|$
of mismatch between the corresponding components of the network. Monte-Carlo
simulations are used for this analysis whereby all magnitudes and
phases of all S-parameters and the LNA $\Gamma_{\text{opt}}$s are
randomly varied within 5\% and 1\% of their original values, thereby
creating maximum variations of 10\% and 2\%, respectively, between
the corresponding components.

\begin{figure}
\begin{centering}
\subfloat[]{\begin{centering}
\includegraphics[width=0.5\columnwidth]{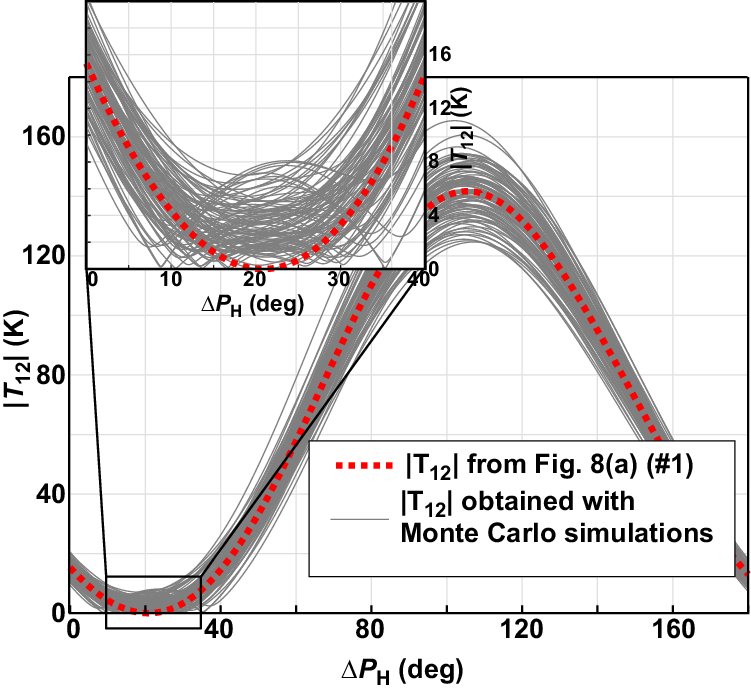}
\par\end{centering}
}\subfloat[]{\begin{centering}
\includegraphics[width=0.5\columnwidth]{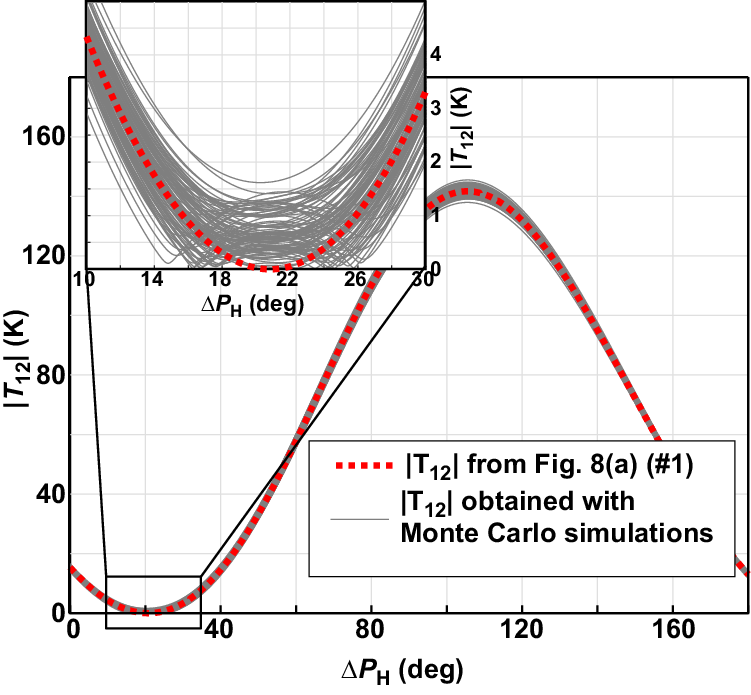}
\par\end{centering}
}
\par\end{centering}
\caption{\label{fig:MC-sims}Monte-Carlo simulation results of $\left|T_{\text{12}}\right|$
for the system configured as with the matching network realizing Location
\#1 of $S_{\text{L,}11}$ and $\Gamma_{\text{opt}}$ in Fig. \ref{fig:Wide-Tij-notch}(b).
In the Monte-Carlo simulations, all S-parameters and $\Gamma_{\text{opt}}$s
are varied so that their maximum variations for each pair of the corresponding
network components were limited to (a) 10\% and (b) 2\%.}
\end{figure}

\begin{figure}
\begin{centering}
\subfloat[10\% variations]{\begin{centering}
\includegraphics[width=0.5\columnwidth]{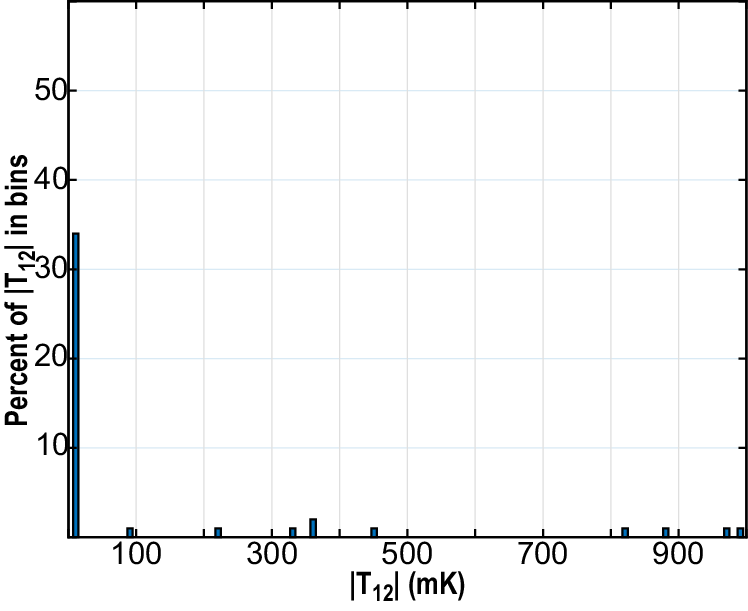}
\par\end{centering}
}\subfloat[2\% variations]{\begin{centering}
\includegraphics[width=0.5\columnwidth]{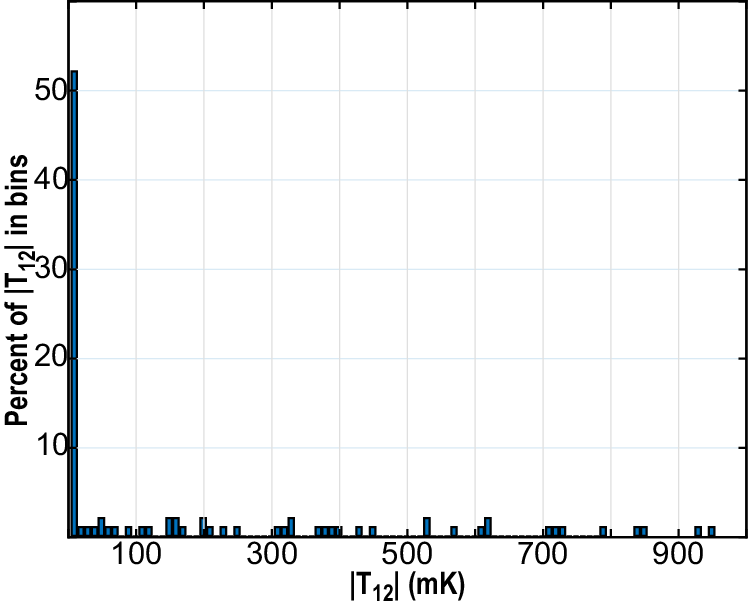}
\par\end{centering}
}
\par\end{centering}
\caption{\label{fig:Histogram}Histogram of $\min\left\{ \left|T_{12}\right|\right\} $
distribution obtained with 100-iteration Monte-Carlo simulations with
(a) 10\% and (b) 2\% variations between S-parameters and the LNA $\Gamma_{\text{opt}}$s
of corresponding network components. In these simulations, $\Delta P_{\text{H}1}$
and $\Delta P_{\text{H}2}$ were independently swept. Each bin is
10~mK wide. Only bins below 1~K are shown.}
\end{figure}

For these simulations, the network realizing Location \#1 for $\Gamma_{\text{opt}}$
and $S_{\text{L,}11}$ in Fig. \ref{fig:Wide-Tij-notch}(b) was selected.
The outcome of these simulations in Fig. \ref{fig:MC-sims} 
show that even the 2\% variations in S-parameters and $\Gamma_{\text{opt}}$s
may result in the minimum of $\left|T_{12}\right|$ exceeding the
10s to 100s~mK range required to discern the perturbations in the
Cosmic Microwave Background. Although an analysis of the appropriate
tuning is not the intent of this work, additional simulations were
still performed to investigate the likelihood of tuning the minima
in the $\left|T_{12}\right|$ curves below 10~mK. In these simulations,
the phase shifter phases $\Delta P_{\text{H},1}$ and $\Delta P_{\text{H},2}$
were tuned independently. Fig. \ref{fig:Histogram} 
shows the results of such analysis, where, for the case of the 10\%
variations, 34\% of simulations returned minima in $\left|T_{12}\right|$
below 10~mK. This percentage increases to 52\% when component mismatches
are limited to 2\%. These simulations suggest that two independently
adjustable phase shifters and pre-selected components may be sufficient
to null mutual coherence even when non-identical components are used.

\subsection{Wideband interferometer}

We next return to identical components and investigate the possibility
of developing a wideband interferometer, such as, for example, would
be used in cosmology instruments, by simulating the network in Fig.
\ref{fig:Block-diagram} in which the S-parameters of the antenna
arrays are taken from calculations presented in \cite{SutinjoURI2020},
Mini-Circuits PMA2-43LN+ (biased with 60~mA current) is used for
the LNA, and Mini-Circuits JSPQW-100A+ is used for the hybrid. This
LNA was selected due to the availability of S-parameters and noise
parameters in its datasheet. Still, the noise parameters are only
given for frequencies above 800~MHz and the gain is low in the 50-to-100-MHz
range. To rectify this, the noise parameters were linearly extrapolated
to low frequencies as: $T_{\text{min}}\approx T_{0}\times0.06f_{\text{GHz}}$,
$N\approx0.34-0.3\times f_{\text{GHz}}$, $Y_{\text{opt}}\approx\left(0.01f_{\text{GHz}}+0.004\right)-j0.005f_{\text{GHz}}$,
where $Y_{\text{opt}}=Z_{0}^{-1}\left(1-\Gamma_{\text{opt}}\right)/\left(1+\Gamma_{\text{opt}}\right)$.
All temperatures in $\mathbf{T}_{\text{PASS}}$ and $T_{\text{L}}$
are set to $290\,\text{K}$. The S-parameters were used directly,
as the low gain does not affect the results since the interferometer
is assumed noiseless. In a practical implementation, an LNA developed
specifically for this application would have sufficient gain to minimize
the impact of the interferometer noise on the system performance.

The selected array is wideband and consists of two closely spaced
short antenna dipoles over a ground plane for smooth response. The
close spacing enables a response to the isotropic noise temperature
component of the surrounding medium. As the result, the antennas are
highly reflective. The $S_{\text{L},11}$ and $\Gamma_{\text{opt}}$
of the selected LNA are also very reflective within the 50-to-100-MHz
band.  Numerical calculation results in Fig. \ref{fig:CrossCorr-nothing-ideal}
illustrate that large $\left|S_{\text{L},11}\right|$ and $\left|\Gamma_{\text{opt}}\right|$
are acceptable for minimizing $\left|T_{12}\right|$, although $\left|S_{\text{L},11}\right|\approx0$
is preferred for the smooth trough in $\left|T_{12}\right|$ (see
Fig. \ref{fig:Sim-results-cross-corr}(d)) and $\Gamma_{\text{opt}}=S_{\text{H},11}$
results in the lowest $T_{\text{rec}}$. 

\begin{figure}
\begin{centering}
\subfloat[]{\includegraphics[width=0.45\columnwidth]{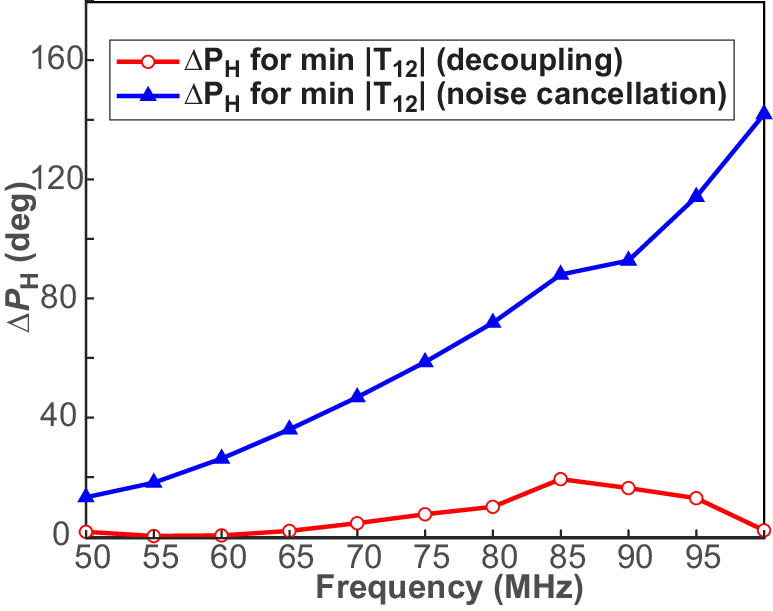}

}\subfloat[]{\includegraphics[width=0.45\columnwidth]{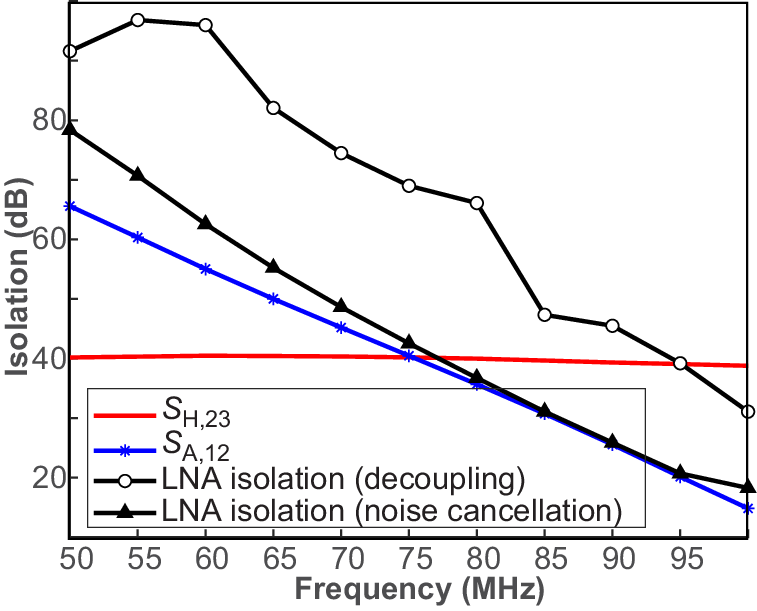}

}
\par\end{centering}
\begin{centering}
\subfloat[]{\includegraphics[width=0.45\columnwidth]{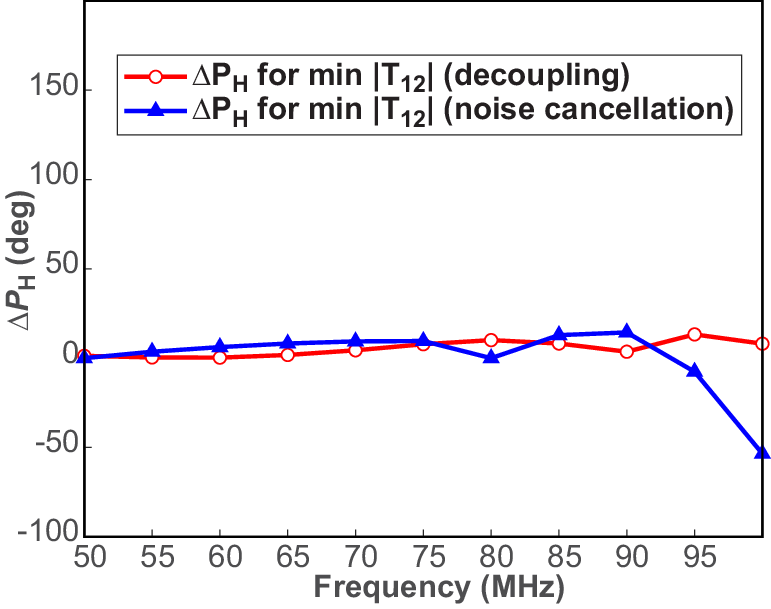}

}
\par\end{centering}
\centering{}\caption{\label{fig:NothingIdeal}a) Extra phase from the phase shifters for
minimum $\left|T_{12}\right|$ versus frequency; b) Simulated isolation
between antennas ($S_{\text{A},12}$) and between the coupled ports
of the hybrid ($S_{\text{H,23}}$), as well as the effective isolation
of the LNAs realized with the network when $\Delta P_{\text{H}}$
as in (a); and c) Extra phase from the phase shifters for minimum
$\left|T_{12}\right|$ versus frequency when $S_{\text{L},11}$ and
$\Gamma_{\text{opt}}$ are near zero.}
\end{figure}

Fig. \ref{fig:NothingIdeal} 
shows the results of the numerical analysis in which two identical
ideal phase shifters (i.e. $\Delta P_{\text{H},1}=\Delta P_{\text{H},2}=\Delta P_{\text{H}}$)
are inserted at the $90^{\circ}$ terminals of the hybrids. The two
phase-shifter phase curves in Fig. \ref{fig:NothingIdeal}(a) correspond
to the conditions of noise cancellation and decoupling. The required
phase shifts to null mutual coherence exhibit monotonic behavior (see
Fig. \ref{fig:NothingIdeal}(a)). At low frequencies, the coupling
is dominated by the hybrid (Fig. \ref{fig:NothingIdeal}(b)). The
decoupling phase-shifter phase is able to significantly increase the
isolation between the LNAs beyond what is afforded by the antenna
array itself. At higher frequencies, the array coupling dominates.
The network is still able to null mutual coherence but at the cost
of larger changes in the required phase shifter phase and higher sensitivity
to the phase.  When $S_{\text{L},11}$ and $\Gamma_{\text{opt}}$
are reduced to nearly zero (1\% of their original values), the optimum
$\Delta P_{\text{H}}$ for the two conditions come close to each other
and remain near-zero over most of the bandwidth, as shown in Fig.
\ref{fig:NothingIdeal}(c), indicating that with careful engineering
of the LNA, wideband nulling in mutual coherence and coincidental
nulls for less sensitivity to hybrid phase is likely possible.

\section{\label{sec:Discussions}Discussions}

The inclusion of a replica array and $90^{\circ}$ hybrids is shown
to provide a means of nulling the mutual coherence at the output of
a two-element interferometer. Two conditions for nulling are identified:
decoupling and noise cancellation. Matching networks can be used to
reduce the sensitivity of mutual coherence nulls to hybrid phase.
The noise temperature of the interferometer is dominated by the ambient
temperature of the replica array unless the latter is cooled. However,
when nulling is achieved, the two-element interferometer becomes sensitive
to even low-level correlated inputs such as isotropic noise temperature
components of the surrounding medium.

For cosmology experiments in the 50-to-100-MHz range, the sky noise
temperature $T_{\text{sky}}\approx60\lambda^{2.55}$ \cite{nijboer2013lofar},
which corresponds to $T_{\text{sky}}$ between 5800 and 1000\,K.
These values are larger than $T_{\text{rec}}$, the beam-equivalent
receiver noise temperature of the interferometer, even without cooling
(see Fig. \ref{fig:Trec-nothing-ideal}), and are much larger in the
limiting case when the replica array and the hybrids are cooled to
nearly 0 K (see Figs. \ref{fig:Sim-results-LNA-coupling}(a)-(c)).
Note that for cosmology experiments, $T_{\text{rec}}$ affects the
integration time, and therefore keeping it below $T_{\text{sky}}$
would result in minimum integration penalty. As a result, the proposed
mutual-coupling canceler is of potential interest for radio cosmology.

\section{\label{sec:Conclusion}Conclusion}

This paper investigated the use of a previously-proposed self-interference
canceler on the noise performance of two-element phased arrays and
interferometers. In this work, the canceler acts as a mutual-coupling
canceler. When used in phased arrays, the canceler makes LNA noise
matching independent of the beamformer coefficients but increases
the beam-equivalent noise temperature by the physical temperature
of the canceler. In an interferometric applications, the addition
of the canceler nulls mutual coherence in the output of the interferometer
while realizing gain in response to correlated inputs to the antennas.
It was shown that the accuracy of detecting the noise temperature
of a uniform sky, $T_{\text{sky}}$, with the proposed network relies
on the accuracy of measuring its physical temperature. With non-ideal,
realistic, and wideband hybrid and antenna models, the low input reflection
coefficient and $\text{\ensuremath{\Gamma_{\text{opt}}}}$ of LNAs
realize two nulls in the mutual coherence that are nearly coincidental
even without additional matching networks or means of tuning the hybrid
phases. However, when there are practical mismatches in network components,
Monte-Carlo analysis demonstrated that to lower mutual coherence,
there is a need for adjusting phases of each hybrid independently.

The analysis in this work employed a general framework, which can
also be used for a large variety of other networks, such as an implementation-ready
network of the two-element interferometer that includes tuning networks,
additional amplifiers, any cables, and a non-ideal interferometer.

Future work will include thorough analysis of the effect of mismatches
between the system components in a wideband system. The proposed framework
for analyzing noise will be used to determine the advantageous combinations
of the component S-parameters that are both realizable and able to
create wideband nulls in mutual coherence. This will lead to identifying
the minimum sufficient set of variable/tunable network parameters
required to null mutual coherence even with not-exactly-identical
components and to ultimately enable the experimental validation of
the results of this work.

\bibliographystyle{IEEEtran}
\bibliography{bibliography}

\begin{IEEEbiography}[{\begin{minipage}[t][1.25in]{1in}%
\includegraphics[width=1in,height=1.25in]{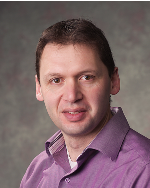}%
\end{minipage}}]{Leonid Belostotski}
 (Senior Member, IEEE) received the B.Sc. and M.Sc. degrees in electrical
engineering from the University of Alberta, Edmonton, AB, Canada,
in 1997 and 2000, respectively, and the Ph.D. degree from the University
of Calgary, Calgary, AB, Canada, in 2007. He was a RF Engineer with
Murandi Communications Ltd., Calgary, from 2001 to 2004. He is currently
a Professor with the University of Calgary and the Canada Research
Chair in high-sensitivity radiometers and receivers. His current research
interests include RF and mixed-signal ICs, high-sensitivity receiver
systems, antenna arrays, and terahertz systems. He was a recipient
of the Outstanding Student Designer Award from Analog Devices Inc.,
in 2007, and the IEEE Microwave Theory and Techniques-11 Contest on
Creativity and Originality in Microwave Measurements, in 2008. He
serves as the IEEE Southern Alberta Solid-State, Circuits and Circuits
and Systems Chapter\textquoteright s Chair. He served as an Associate
Editor for the IEEE TRANSACTIONS ON INSTRUMENTATION AND MEASUREMENT.
He is the Editor-in-Chief for the IEEE SOLID-STATE CIRCUITS MAGAZINE.
\end{IEEEbiography}

\begin{IEEEbiography}[{\begin{minipage}[t][1.25in]{1in}%
\includegraphics[width=1in,height=1.25in]{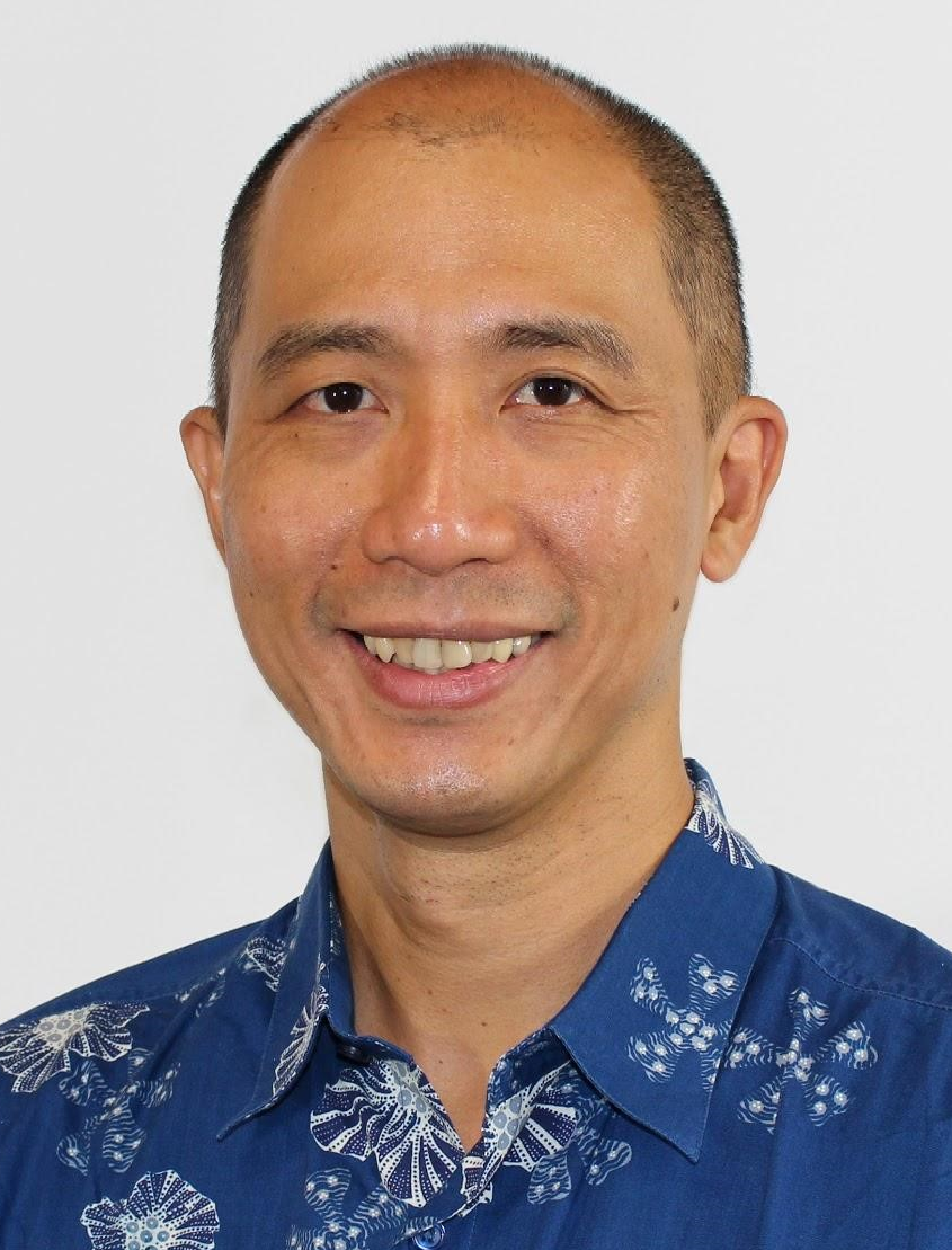}%
\end{minipage}}]{Adrian T. Sutinjo}
  (Senior Member 2015) received the B.S.E.E. degree from Iowa State
University, Ames, IA, USA, in 1995, the M.S.E.E. degree from the Missouri
University of Science and Technology, Rolla, MO, USA, in 1997, and
the Ph.D. degree in electrical engineering from the University of
Calgary, Calgary, AB, Canada, in 2009. From 1997 to 2004, he was an
RF Engineer with Motorola, Schaumburg, IL, USA, and with Murandi Communications
Ltd., Calgary. He is currently a Senior Lecturer with the School of
Electrical Engineering, Computing and Mathematics, Curtin University,
Perth, WA, Australia. He has been with the International Centre for
Radio Astronomy Research/Curtin Institute of Radio Astronomy, Curtin
University, Bentley, WA, Australia, since 2012. His current research
interests include antennas, RF and microwave engineering, electromagnetics,
and radio astronomy engineering.
\end{IEEEbiography}

\begin{IEEEbiography}[{\begin{minipage}[t][1.25in]{1in}%
\includegraphics[width=1in,height=1.25in]{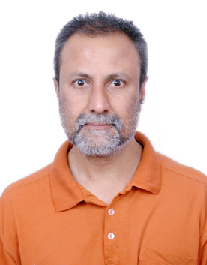}%
\end{minipage}}]{Ravi Subrahmanyan}
  received the B.Tech degree in Electrical Engineering from the
Indian Institute of Technology, Madras, India, in 1983 and the Ph.D.
degree in astronomy from the Physics department of the Indian Institute
of Science, Bangalore, India, in 1990. He is with Space \& Astronomy,
CSIRO, Australia.
\end{IEEEbiography}

\begin{IEEEbiography}[{\begin{minipage}[t][1.25in]{1in}%
\includegraphics[width=1in,height=1.25in]{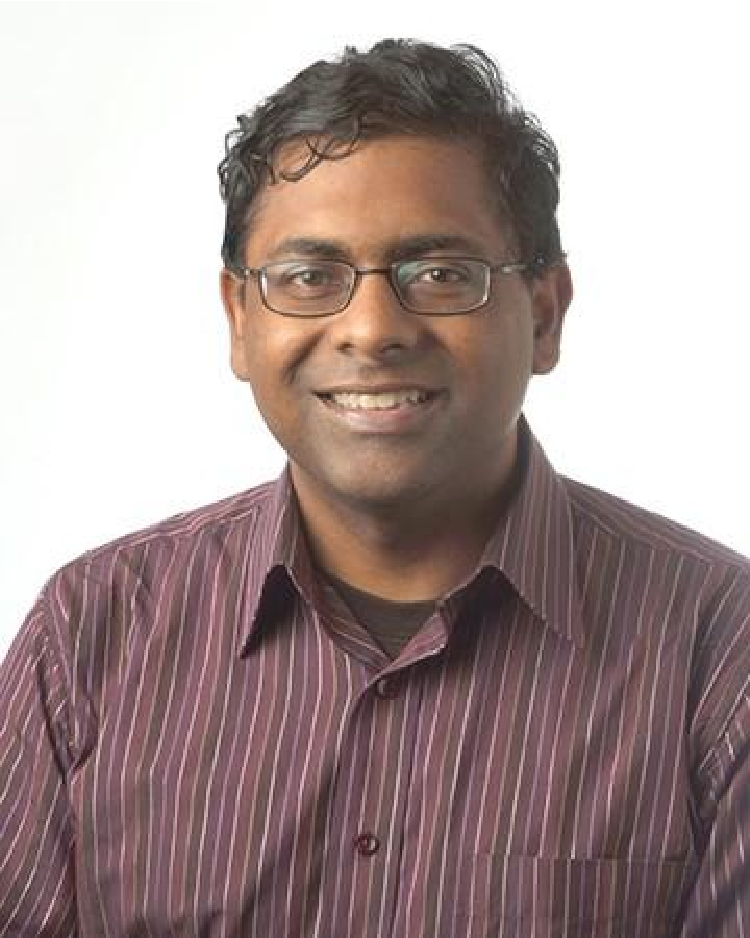}%
\end{minipage}}]{Soumyajit Mandal}
  (S'01-{}-M'09-{}-SM'14) received the B.Tech. degree from the Indian
Institute of Technology (IIT) Kharagpur, India, in 2002, and the S.M.
and Ph.D. degrees in electrical engineering from the Massachusetts
Institute of Technology (MIT), Cambridge, MA, USA, in 2004 and 2009,
respectively.

He was a Research Scientist with Schlumberger-Doll Research, Cambridge
(2010 - 2014), an Assistant Professor with the Department of Electrical
Engineering and Computer Science, Case Western Reserve University,
Cleveland, OH, USA (2014 -- 2019), and an Associate Professor with
the Department of Electrical and Computer Engineering, University
of Florida, Gainesville, FL, USA (2019 -- 2021). He is currently
a Senior Engineer in the Instrumentation Division at Brookhaven National
Laboratory, Upton, NY, USA. He has over 150 publications in peer-reviewed
journals and conferences and has been awarded 26 patents. His research
interests include analog and biological computation, magnetic resonance
sensors, low-power analog and RF circuits, and precision instrumentation
for various biomedical and sensor interface applications. He was a
recipient of the President of India Gold Medal in 2002, the MIT Microsystems
Technology Laboratories (MTL) Doctoral Dissertation Award in 2009,
the T. Keith Glennan Fellowship in 2016, and the IIT Kharagpur Young
Alumni Achiever Award in 2018.
\end{IEEEbiography}

\begin{IEEEbiography}[{\begin{minipage}[t][1.25in]{1in}%
\includegraphics[width=1in,height=1.25in]{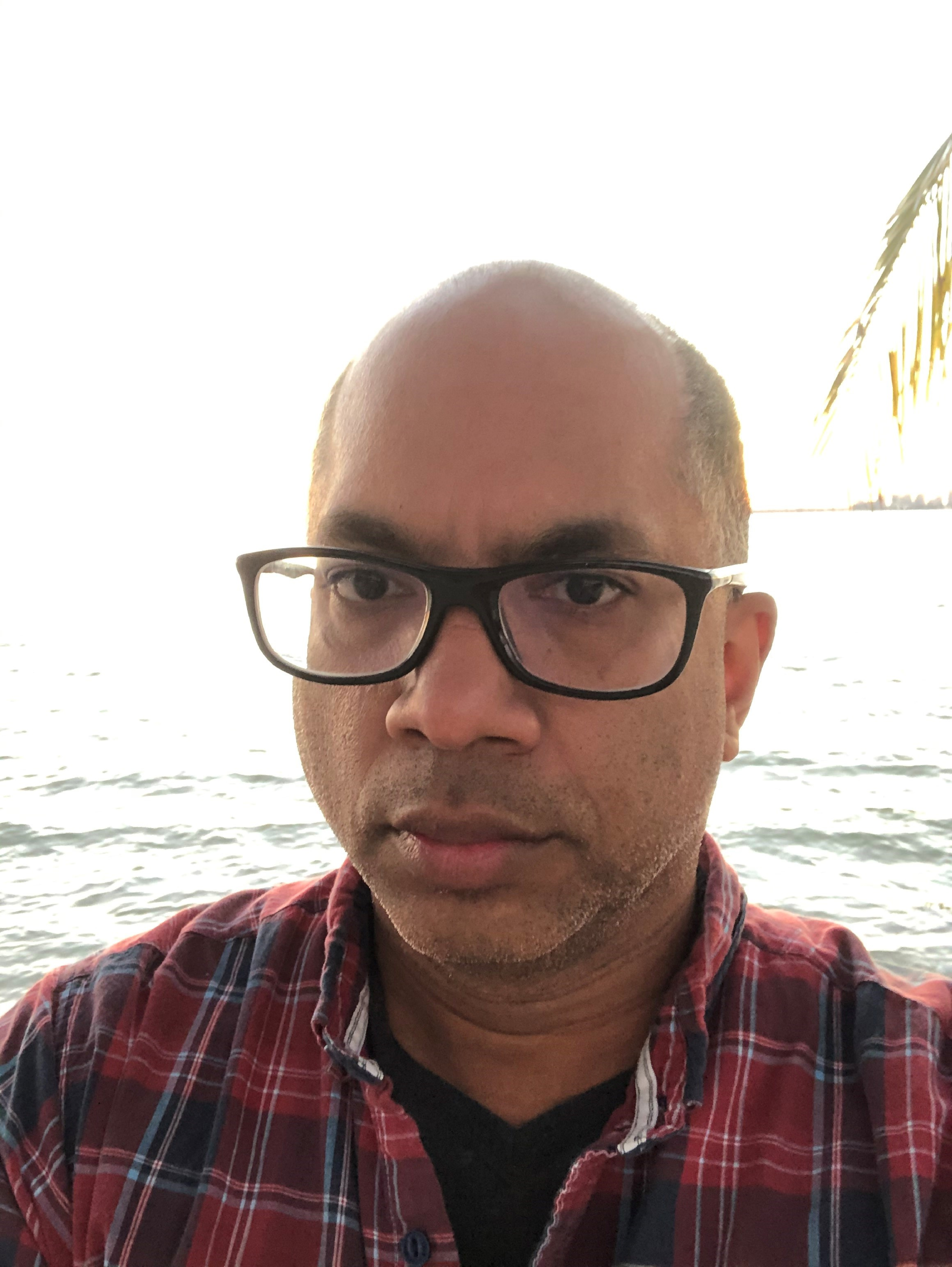}%
\end{minipage}}]{Arjuna Madanayake}
  is an Associate Professor at the Dept. of Electrical and Computer
Engineering (ECE) at Florida International University (FIU), Miami,
Florida. He as awarded the Ph.D. and M.Sc. degrees both in Electrical
Engineering from the University of Calgary, AB,Canada, in 2008 and
2004, respectively. He completed the B.Sc. {[}Eng{]} in Electronic
and Telecommunication Engineering- passing with First Class Honors
- from the University of Moratuwa, Sri Lanka, in 2002. Dr. Madanayake
has research interests spanning signal processing, wireless communications,
analog and digital circuits, RF systems, antenna arrays, FPGA systems,
radar and electronic warfare, fast algorithms and VLSI, analog computing,
radio astronomy instrumentation, 5G/6G mm-wave systems, advanced spectrum,
blockchain applications for spectrum, RF-machine learning, AI accelerators
for 6G, and software defined radio (SDR). He leads the RF, Analog
and Digital (RAND) Lab for Advanced Signal Processing Circuits (ASPC)
where his research has been recently supported by multiple grants
from NSF, ONR, NASA, NIH and DARPA.
\end{IEEEbiography}

\end{document}